\documentclass[%
 aip,
floatfix,
amsmath,
amssymb,
preprint,%
]{revtex4-1}

\usepackage{graphicx}
\usepackage{dcolumn}
\usepackage{bm}

\usepackage[utf8]{inputenc}
\usepackage[T1]{fontenc}
\usepackage{mathptmx}
\usepackage{etoolbox}

\usepackage{amsmath}
\usepackage{amsfonts}
\usepackage{physics}
\usepackage{bm}
\usepackage{subcaption}
\usepackage{makecell}
\usepackage{ifthen}
\usepackage{nomencl}
\usepackage{hyperref}

\newcommand{\unn}[1][]{%
	\ifthenelse{\equal{#1}{}}{\bm{\hat{u}}(\bm{\theta}, \bm{x}, t)}{\bm{\hat{u}}(\bm{\theta}, \bm{x}_{#1}, t_{#1})}
}
\newcommand{\unnInit}[1][]{%
	\ifthenelse{\equal{#1}{}}{\bm{\hat{u}}(\bm{\theta}, \bm{x}, 0)}{\bm{\hat{u}}(\bm{\theta}, \bm{x}_{#1},0)}%
}
\newcommand{\unnStat}[1][]{%
	\ifthenelse{\equal{#1}{}}{\bm{\hat{u}}(\bm{\theta}, \bm{x})}{\bm{\hat{u}}(\bm{\theta}, \bm{x}_{#1})}%
}
\newcommand{\unnPar}[1][]{%
	\ifthenelse{\equal{#1}{}}{\bm{\hat{u}}(\bm{\theta}, \bm{x}, t)}{\bm{\hat{u}}(\bm{\theta}, \bm{x}, t, {#1})}%
}
\newcommand{\unnStatPar}[1][]{%
	\ifthenelse{\equal{#1}{}}{\bm{\hat{u}}(\bm{\theta}, \bm{x})}{\bm{\hat{u}}(\bm{\theta}, \bm{x}, {#1})}%
}

\makeatletter
\def\@email#1#2{%
 \endgroup
 \patchcmd{\titleblock@produce}
  {\frontmatter@RRAPformat}
  {\frontmatter@RRAPformat{\produce@RRAP{*#1\href{mailto:#2}{#2}}}\frontmatter@RRAPformat}
  {}{}
}%
\makeatother
\begin{document}


\title[]{Physics-Informed Neural Networks for Transonic Flows around an Airfoil}
\author{S. Wassing}
\affiliation{German Aerospace Center - Institute for Aerodynamics and Flow Technology \\
             Center for Computer Applications in Aerospace Science and Engineering}
\email{simon.wassing@dlr.de}
\author{S. Langer}%
\affiliation{German Aerospace Center - Institute for Aerodynamics and Flow Technology \\
             Center for Computer Applications in Aerospace Science and Engineering}
\author{P. Bekemeyer}
\affiliation{German Aerospace Center - Institute for Aerodynamics and Flow Technology \\
             Center for Computer Applications in Aerospace Science and Engineering}

\date{\today}

\begin{abstract}
Physics-informed neural networks have gained popularity as a deep-learning based parametric partial differential equation solver. Especially for engineering applications, this approach is promising because a single neural network could substitute many classical simulations in multi-query scenarios. Only recently, researchers have successfully solved subsonic flows around airfoils with physics-informed neural networks by utilizing mesh transformations to precondition the training. However, compressible flows in the transonic regime could not be accurately approximated due to shock waves resulting in local discontinuities. In this article, we propose techniques to successfully approximate solutions of the compressible Euler equations for sub- and transonic flows with physics-informed neural networks. Inspired by classical numerical algorithms for solving conservation laws, the presented method locally introduces artificial dissipation to stabilize shock waves. We compare different viscosity variants such as scalar- and matrix-valued artificial viscosity, and validate the method at transonic flow conditions for an airfoil, obtaining good agreement with finite-volume simulations. Finally, the suitability for parametric problems is showcased by approximating transonic solutions at varying angles of attack with a single network. The presented work enables the application of parametric neural network based solvers to a new class of industrially relevant flow conditions in aerodynamics and beyond. 
\end{abstract}

\maketitle

\section*{Nomenclature}
\subsection*{List of Symbols}
\begin{tabular}{cl}
	$\bm{W}$ & vector of conserved variables\\
    $\bm{F}_x$ & flux vector in x direction\\
	$\bm{F}_y$ & flux vector in y direction\\
    $\rho$ & density\\
	$u$ & velocity in x-direction\\
	$v$ & velocity in y-direction\\
	$E$ & total specific energy\\
	$p$ & pressure\\
    $t$ & time\\
    $\kappa$ & ratio of specific heats\\
    $x$ & first Cartesian coordinate in physical domain\\
	$y$ & second Cartesian coordinate in physical domain\\
    $\Omega$ & physical domain\\
    $\Omega_\mathrm{h}$ & grid in $\Omega$\\
    $\xi$ & first curvilinear coordinate\\
    $\eta$ & second curvilinear coordinate\\
    $\Sigma$ & computational domain\\
    $\Sigma_\mathrm{h}$ & grid in $\Sigma$\\
    $J$ & Jacobian of mesh transformation\\
    $\mathcal{L}$ & loss function\\
    $\mathcal{L_\mathrm{res}}$ & residual loss term\\
    $\mathcal{L_\mathrm{bdr}}$ & boundary loss term\\
    $\lambda_\mathrm{res}$ & residual loss term weight\\
    $c$ & speed of sound\\
    $M$ & Mach number
\end{tabular}
\newpage
\begin{tabular}{cl}
    $\bm{w}_\infty$ & vector of primitive variables in far-field\\
    $\rho_\infty$ & density in far-field\\
    $u_\infty$ & velocity in x-direction in far-field\\
    $v_\infty$ & velocity in y-direction in far-field\\
    $p_\infty$ & pressure in far-field\\
    $M_\infty$ & Mach number in far-field\\
    $\alpha$ & angle of attack\\
    $\bm{q}$ & velocity vector\\
    $MAE$ & mean absolute error \\
    $RMAE$ & relative mean absolute error\\
    $R_2$ & coefficient of determination / $R_2$-score\\
    $\partial\Sigma_\infty$ & upper boundary of computational domain\\
    $\partial\Sigma_\mathrm{af}$ & lower boundary of computational domain\\
    $\partial\Sigma_\mathrm{l}$ & left boundary of computational domain\\
    $\partial\Sigma_\mathrm{r}$ & right boundary of computational domain\\
    $N$ & number of grid points\\
    $N_\xi$ & number of points on upper and lower boundary of c. domain\\
    $N_\eta$ & number of points of left and right boundary of c. domain\\
    $\hat{\bm{u}}$ & neural network output vector\\
    $\mu_x, \mu_y$ & artificial viscosity in x and y direction\\
    $\nu$ & artificial viscosity factor\\
    $A$ & flux jacobian in x-direction\\
    $B$ & flux jacobian in y-direction\\
    $s$ & sensor function (combined)\\
    $s_\mathrm{shock}$ & shock sensor\\
    $k^{(0)}_{\mathrm{shock}}$ & activation threshold of shock sensor\\ 
    $k^{(1)}_{\mathrm{shock}}$ & activation steepness of shock sensor\\
    $s_\mathrm{stag}$ & stagnation point sensor\\
    $k^{(0)}_{\mathrm{stag}}$  & activation threshold of stagnation sensor\\
    $k^{(1)}_{\mathrm{stag}}$  & activation steepness of stagnation sensor
\end{tabular}	

\begin{tabular}{cl}
    $\bm{r}$ & general input vector to neural network or Fourier embedding\\
    $f^k$ & layer function in neural network\\
    $\sigma^k$ & activation function for $k$-th network layer\\
    $w^k$ & weight matrix in $k$-th network layer\\
    $D$ & number of hidden network layers\\
    $N_k$ & number of neurons in $k$-th network layer\\
    $\omega_\sigma^k$ & trainable parameter in adaptive activation in $k$-th layer\\
    $g$ & magnitude of factorized weights\\
    $\bm{z}$ & direction vector of factorized weights\\
    $\hat{\bm{u}}_\mathrm{fe}$ & output vector of Fourier embedding layer\\
    $\phi_{i, j}$ & Fourier frequency\\
    $\sigma_\phi$ & standard deviation of normal distribution for sampling $\phi_{i, j}$\\
    $C_p$ & coefficient of pressure\\
    $N_\alpha$ & number of $\alpha$ samples\\
    $N_{\alpha, \mathrm{bound}}$ & number of $\alpha$ values on bounds of parameter range\\
    $p_\mathrm{tot}$ & total pressure\\
    $C_{p_\mathrm{tot}, \mathrm{loss}}$ & total pressure loss
\end{tabular}

\vspace{1.cm}
\subsection*{Abbreviations}
\begin{tabular}{cl}
	PINN & physics-informed neural network\\
	NN & neural network\\
        ADAM & adaptive momentum estimation optimization algorithm\\
        L-BFGS & Limited-memory Broyden-Fletcher-Goldfarb-Shanno algorithm\\
	PDE & partial differential equation\\
	AV & artificial viscosity\\
	CFD & computation fluid dynamics\\ 
        FV & finite volume\\
        NACA & National Advisory Committee for Aeronautics\\
        RANS & Reynolds-averaged Navier-Stokes
\end{tabular}

\clearpage

\section{Introduction}
The solution of partial differential equations (PDEs) is of major importance in many fields of science and engineering, such as aerodynamics. One notable example is the solution of the Reynolds-Averaged Navier-Stokes (RANS) equations, representing the de-facto standard for many industrial aerodynamic applications such as the simulation of flows around transport aircraft. Oftentimes, the RANS equations represent a sensible compromise between accuracy and computational cost, allowing for accurate predictions of aerodynamic performance in a reasonable simulation time. However, to achieve reasonable accuracies for complex three-dimensional geometries at high Reynolds-numbers, the computational cost can still be prohibitive. In addition, one is oftentimes not interested in a single set of the flow conditions, but in several variations of the problem conditions, resulting in variable parameters in the underlying PDE as well as the boundary and initial conditions. Such many-query scenarios are for example common during design exploration and optimization. 
Recently, physics-informed neural networks, introduced by Raissi~et~al.~\cite{Raissi.2017, Raissi.2017b, Raissi.2019}, have emerged as a novel approach for solving boundary value problems. Their ability to handle parametric problems is particularly promising, since a single neural network has to be trained only once to solve a boundary value problem for an entire range of varying parameter conditions~\cite{Wassing.2024, Hennigh.2021, Lorenzen.2024, Cao.2024c, Cao.2025b}, making them highly efficient for multi-query scenarios. The PINN's loss function can even be augmented with supervised terms to solve inverse problems~\cite{Raissi.2019, Mao.2020, Jagtap.2022, Raissi.2020, Mommert.2024}. For example, experimental measurements can be directly incorporated into the training process of the model.\par
To fully realize the potential of parametric PINNs, it is essential to establish the basics for solving challenging PDEs with PINNs. By doing so, researchers can unlock new possibilities for faster design exploration and optimization in various fields, including aerodynamics. Looking ahead, the integration of PINNs with emerging computing architectures, such as quantum computers, may offer even greater opportunities for acceleration. Recent advances have shown that PINNs can be implemented on quantum computers using Physics-Informed Quantum Circuits (PIQCs)~\cite{Kyriienko.2021, Siegl.2024}, which could potentially lead to breakthroughs in efficiently finding solutions of complex flows. However, for now, the primary focus remains on developing the fundamentals of PINNs and their application to parametric problems.
While the original formulation of PINNs has been successful in solving many academic problems governed by PDEs of various kinds, the training has shown to be challenging, especially when trying to apply PINNs to complex problems.\par
In this article, we are interested in solving boundary-value problems for the compressible Euler equations using PINNs. The Euler equations are, for example, obtained, when the viscous terms in the compressible RANS equations are neglected. Hence, the Euler equations model inviscid compressible flows, which are of particular interest in aerospace engineering, since they describe many phenomena of relevance, such as shock waves. Because of these compressibility effects the application of PINNs is not straightforward. In order to correctly resolve discontinuities which can occur in the solutions, adjustments are necessary. A robust and accurate implementation of these convection equations is an essential component in order to be able to tackle the compressible RANS equations in a further step.\par
In general, many extensions to the initial PINN methodology to improve the prediction accuracy have been proposed. Imbalances between the gradients of different loss terms have been identified as a possible reason for poor convergence of PINNs~\cite{Wang.2021}. To address this issue, adaptive weighting strategies for automatically balancing the loss terms during training, have been developed~\cite{Wang.2021, Jin.2021, Maddu.2022, Wang.2023}. Modifications to the trainable layers of PINNs, based on adaptive activation functions or weight normalization seem to be able to improve the convergence~\cite{Jagtap.2020c, Jagtap.2020b, Salimans.2016}. Wang~et~al.~\cite{Wang.2023} provided an overview of some of the most crucial established techniques. While such general techniques can help to improve and overcome typical limitations of PINNs, the mathematical intricacies of particular PDEs may give rise to additional issues requiring special attention. Here, numerical legacy methods can serve as a paradigm of how such issues can be overcome.\par
With a more specific aerodynamic focus, Cao~et~al.~\cite{Cao.2024b} have shown accurate predictions for subsonic inviscid and steady-state flows around an airfoil by combining PINNs with mesh transformations. This classical methodology is well known in the context of computational fluid dynamics (CFD) (see e.g. \cite[pp. 168-215]{Anderson.1995}) and can, for example, be used to solve PDEs for non-trivial geometries with finite differences. The general idea is to transform a curvilinear grid representing the physical domain into a regular grid in the computational domain. The boundary value problem is then solved on the regular grid in this computational domain. PINNs do not strictly require a regular grid-like point distribution. It was, however, demonstrated that using mesh transformation yields significant improvements in accuracy and convergence speed for typical two-dimensional aerodynamic flows around an airfoil. High point-density areas in the physical domain are stretched in the computational domain (e.g. on the airfoil surface) and low point-density areas in the physical domain are clinched in the computational domain. This mitigates some of the natural multi-scale characteristics of such aerodynamic flows. Near an airfoil, we typically have curvature effects in the flow field on very short length scales. The total domain needs to be comparatively large though (multiple chord lengths) to emulate an infinite domain where all flow quantities recede to free stream conditions. In addition, outgoing disturbances may be reflected back into the domain by the far-field boundary, impeding convergence~\cite[pp.262-268]{Blazek.2015}. Therefore, numerical legacy methods require domain sizes around 10-100 times the chord length of the airfoil. Multi-scale problems like this have shown to be challenging for PINNs~\cite{Wang.2021b, Moseley.2023} and the reformulation of the problem in terms of a square computational domain with equalized length scales leads to a loss landscape which is evidently far easier to navigate during training. The PINN prediction matches well with finite volume reference results even on coarse grids. In the following paper, Cao~et~al.~\cite{Cao.2024c} demonstrated how the PINN with mesh transformations can be used to predict the flow around a fully parameterized airfoil shape, highlighting the potential for geometry optimization. The mesh-transformation based PINNs have since also been adapted to solve other parametric PDEs for flows around airfoils, including laminar flows, governed by the incompressible Navier-Stokes equations~\cite{Cao.02.01.2025}, and even the incompressible RANS equations~\cite{Cao.2025b}, clearly demonstrating the major potential of parametric neural network based PDE solvers in aerodynamics. However, the authors also show that for transonic flows, the presented method is unable to accurately approximate the expected normal shock wave~\cite{Cao.2024b}. This is a major limitation for engineering applications in aeronautics, since, for the majority of the time, today's airplane's cruise speed is at transonic conditions.\par

Here, we combine the mesh transformation methodology with the concept of artificial viscosity (AV) and, for the first time, successfully simulate transonic flows around an airfoil with PINNs. In the context of finite volume methods, it is well known that a stable integration scheme for hyperbolic conservation laws requires some form of numerical viscosity. The viscosity is typically introduced explicitly in central difference schemes with additional viscous terms or implicitly included in upwinding schemes \cite[pp. 89-113]{Blazek.2015}. From a mathematical perspective, the use of AV can be motivated by the fact that physically reasonable solutions, so called entropy solutions, of systems of conservation laws are defined as the limit of vanishing viscosity~\cite{Evans.2008}.\par
In the context of PINNs, Fuks~et~al.~\cite{Fuks.2020} first noticed that additional viscous terms are critical to accurately approximate solutions to one-dimensional hyperbolic PDEs, when discontinuities are present. Since then, various works have made use of AV for stabilizing shock waves in hyperbolic conservation laws~\cite{Patel.2022, Coutinho.2023, Wagenaar.2023, Wassing.2024}.\par Besides AV, other methodologies for improving the shock capturing capabilities of PINNs have been proposed~\cite{Mao.2020,Patel.2022, Liu.2024, Jagtap.2022, Jin.2024, Ryck.2024, Chaumet.2024, Lorin.2024}. Notable examples include, Mao~et~al.~\cite{Mao.2020} who use an increased number of collocation points in the neighborhood of shocks to improve the prediction accuracy. Liu~et~al.~\cite{Liu.2024} introduce a physics-dependent weight into the equations, essentially decreasing the impact of the discontinuous regions on the PDE based loss terms. In these regions they instead impose the Rankine-Hugoniot conditions ensuring that the flux across the shock is conserved. Jagtap~et~al.~\cite{Jagtap.2022} use an additional loss term penalizing disagreements with an entropy condition, to enforce physically reasonable solutions. They apply this method to inverse supersonic flow problems. Overall, most of these alternative have so far been either limited to scalar conservation laws, the one-dimensional Euler equations, trivial geometries or inverse problems. Here we demonstrate that AV is a viable candidate for capturing shock waves with PINNs without supervised losses, even when dealing with the Euler equations on non-trivial geometries in two dimensions.\par
The main contributions of this work are as follows: We introduce a novel PINN model, specifically designed for compressible sub- and transonic flows. It uses the concept of artificial viscosity (AV) combined with mesh transformations. Inspired by classical numerical methods, the model identifies normal shock waves, based on a newly developed analytical sensor function. We also introduce three AV variants, in analogy to classical scalar-AV~\cite{JAMESON.1981} and matrix-valued-AV~\cite{Turkel.1989, Swanson.1992}. These AV variants are validated on several transonic flow conditions for the NACA~0012 airfoil, providing insides into how a successful AV scheme for PINNs can be constructed. Lastly, the method is applied to a parametric problem with variable angle of attack at a constant Mach number, showcasing the suitability of the proposed model as a parametric solver.\par
The remaining article is structured as follows. Sec.~\ref{sec:methods_overview} starts with a rough overview of the PINN model while Secs.~\ref{sec:methods_euler_equations}--\ref{sec:methods_implementation_details} focus on the different parts of the model in more detail. This includes, Secs.~\ref{sec:methods_artificial_viscosity}-\ref{sec:methods_sensor} which introduce the novel AV methods and the sensor function enabling the approximation of solutions in transonic flow conditions.
In Sec.~\ref{sec:results_baseline} we present the results for non-parametric PINNs and analyze the behavior of the newly developed sensor function on a qualitative and quantitative basis. Sec.~\ref{sec:results_parametric} show the application of the method to a parametric test case with variable angle of attack. In Sec.~\ref{sec:conclusion_and_outlook} we evaluate the results and discuss future implications of this work.
\section{Methods}
\label{sec:methods}
\label{sec:methods_overview}
\begin{figure}
     \centering
     \includegraphics[width=.8\linewidth]{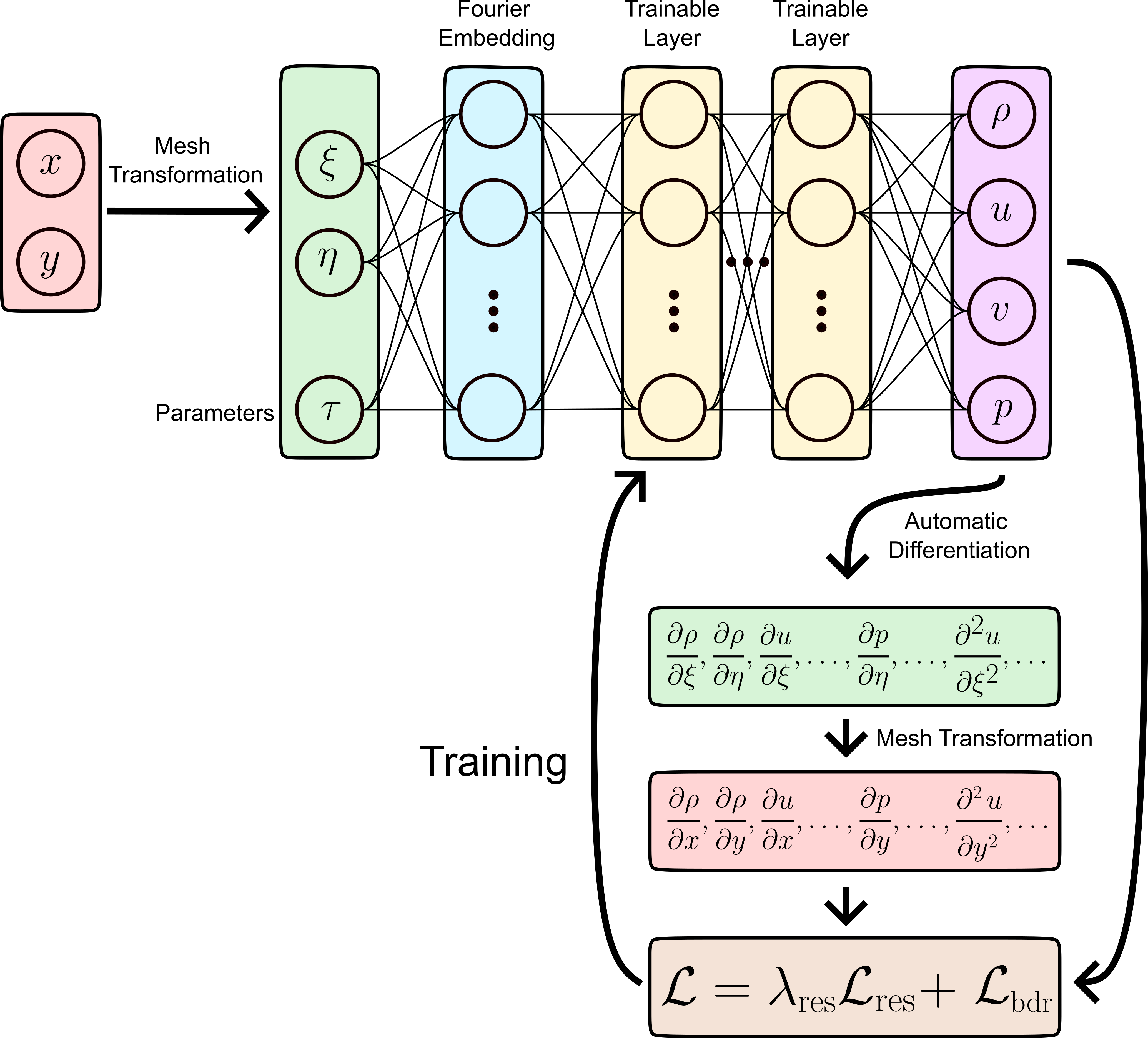}
     \caption{Schematic overview of the used PINN methodology. For each pair of physical coordinates $(x, y)$ their representatives are identified within the computational grid $(\xi, \eta)$. A NN, consisting of a Fourier embedding layer and multiple trainable layers predicts the primitive field variables $(\rho, u, v, p)$. Automatic differentiation is used to calculate the derivatives with respect to $(\xi, \eta)$ and the mesh metrics can be used to reconstruct the derivatives with respect to $(x, y)$. Finally, the model is trained using a loss function consisting of a boundary term $\mathcal{L}_\mathrm{bdr}$ and a residual term $\mathcal{L}_\mathrm{res}$ which includes the developed sensor function. All parts of the model are described in more detail in Secs.~\ref{sec:methods_euler_equations}--\ref{sec:methods_implementation_details}. }
     \label{fig:schematic_method_overview}
 \end{figure}

A schematic overview of the model architecture is shown in Fig.~\ref{fig:schematic_method_overview}. To solve the compressible Euler equations (see Sec.~\ref{sec:methods_euler_equations}), the model is trained at the grid points $(x, y)$ in the physical domain. Each point can be identified with a corresponding point in the computational domain $(\xi, \eta)$ which is passed through a random Fourier-feature embedding layer (Sec.~\ref{sec:methods_fourier_features}). The output is passed through multiple fully connected, trainable layers (Sec.~\ref{sec:methods_trainable_layers}). The final output of the network is the vector of the predicted primitive variables $(\rho, u, v, p)$. Next, we can use automatic differentiation to calculate the derivatives of these predicted variables with respect to the network inputs. Since we do not have an analytic expression for the mesh transformation, we can only calculate the derivatives with respect to the coordinates of the computational domain $\xi$ and $\eta$. However, as explained in Sec.~\ref{sec:methods_mesh_trafo}, the derivatives with respect to the physical coordinates can be reconstructed. These derivatives and the predicted variables themselves are used to calculate the loss function, defined in  Sec.~\ref{sec:methods_loss_function}. Here we modify the original Euler equations by adding an additional viscous term, defined in Sec.\ref{sec:methods_artificial_viscosity}. The term is only active in certain regions of the flow, determined by a sensor function, as described in Sec.~\ref{sec:methods_sensor}. This sensor is crucial for accurately predicting the flow in the transonic regime. Finally, we can use the backpropagation algorithm to optimize the parameters in the trainable layers by minimizing the loss function, thus training the NN to predict the solution of the PDE. Additional parameters of the PDE can be added to the input space of the NN alongside the curvilinear coordinates. The PINN is then trained in a range of parameter values. The trained model can almost instantaneously predict solution in the analyzed parameter range. 

\subsection{Euler Equations}
\label{sec:methods_euler_equations}
We consider boundary-value problems for the 2-D compressible Euler equations,
\begin{equation}
    \begin{aligned}
        \pdv{\bm{W}}{t} + \pdv{\bm{F}_x}{x}+\pdv{\bm{F}_y}{y} &= 0,\\
        \bm{W} = 
        \begin{pmatrix}
            \rho \\
            \rho u \\
            \rho v \\
            \rho E
        \end{pmatrix},\,
        \bm{F}_x = 
        \begin{pmatrix}
            \rho u\\
            \rho u^2 + p \\
            \rho u v \\
            \rho u E + p u \\
        \end{pmatrix},\, 
        \bm{F}_y &= 
        \begin{pmatrix}
            \rho v \\
            \rho u v \\
            \rho v^2 +p \\
            \rho v E + p v \\
        \end{pmatrix},\\
    \end{aligned}
    \label{eq:euler}
\end{equation}
with the density $\rho$, the flow velocities in~$x$- and~$y$-direction $u$ and $v$ and the total energy $E$.
The system is closed by the equation of state for an ideal gas
\begin{equation}
    E = \dfrac{1}{\kappa - 1} \frac{p}{\rho} + \frac{u^2 + v^2}{2}.
\end{equation}
Here $\kappa=1.4$ is the ratio of specific heats for air and~$p$ is pressure.
The speed of sound is then given by $c=\sqrt{\kappa p /\rho}$.\par
We are interested in steady-state solutions $\partial \bm{W} / \partial t = 0$. The NN predicts a vector of the primitive variables ($\rho, u, v, p$). Solving the conservative form of the equations, the fluxes $\bm{F}_x$ and $\bm{F}_y$ are continuous in space, even across shock waves, according to the conservation of mass, momentum and energy.\par

\subsection{Mesh-Transformation}
\label{sec:methods_mesh_trafo}
\begin{figure}[tb]
    \centering
    \includegraphics[width=.6\linewidth]{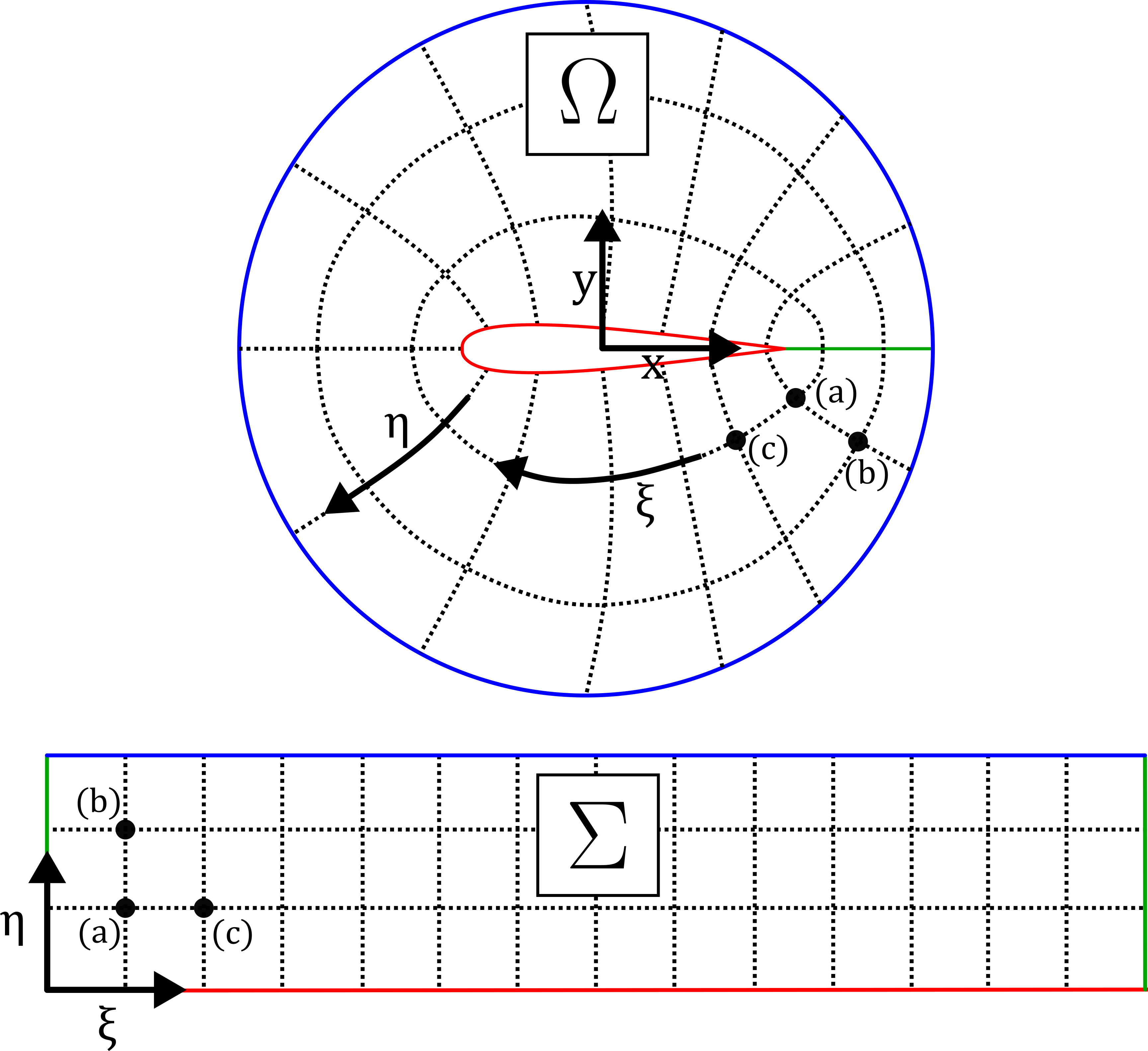}
    \caption{Schematic representation of curvilinear grid (top) in $\Omega$ and corresponding computational grid (bottom) in $\Sigma$. The curvilinear grid wraps around the airfoil. The origin is located at the center. The Cartesian coordinates are $x$ and $y$. The curvilinear coordinates $(\xi, \eta)$ are oriented in tangential and normal direction of the airfoil surface. For the computational grid, the origin is at the lower left. In $\Sigma$, $\xi$ and $\eta$ are the Cartesian coordinates. All grid points of the curvilinear (physical) grid, can be identified with points in the computational grid (e.g. (a), (b), (c)). The outer (blue) and inner (red) domain boundary become the top and bottom boundary in the computational grid. The trailing edge line (green) gets split into the left and right boundary. }
    \label{fig:computational_vs_physical_domain}
\end{figure}

To discretize the boundary-value problem on the \textit{physical domain} $\Omega$, it is mapped into a \textit{computational domain} $\Sigma$ using the invertible and differentiable mapping
\begin{equation}
\mathcal{F}: \Omega \rightarrow \Sigma,\qquad\ (x,y)\mapsto (\xi,\eta).\label{eq:discrete_mesh_trafo}
\end{equation}
Fig.~\ref{fig:computational_vs_physical_domain} illustrates the approach exemplary for a curvilinear grid around an airfoil. In our context the grid is denoted by~$\Omega_h\subset\Omega$, 
\begin{equation}
\Omega_h = \left\{(x_{i,j}, y_{i,j})\in \Omega: i = 1, \dots, N_\xi,\quad  j= 1, \dots, N_\eta\right\}.
\end{equation}
$N_{\xi}$ and~$N_{\eta}$ are the number of points on the grid lines tangential and normal to the airfoil's surface, respectively. The Cartesian grid~$\Sigma_h\subset\Sigma$ in the computational domain is
\begin{equation}
\Sigma_h = \left\{\mathcal{F}(x_{ij},y_{ij})\in\Sigma: (x_{i,j}, y_{i,j})\in \Omega, \quad i = 1, \dots, N_\xi,\,j= 1, \dots, N_\eta\right \}.
\end{equation}
Using the definition~$(\xi_{ij},\eta_{ij}) := \mathcal{F}(x_{ij},y_{ij})$, the function~$ \mathcal{F}$ is constructed in a discrete sense such that
\begin{equation}
     \mathcal{F}(x_{ij},y_{ij}) = (\xi_{ij},\eta_{ij}),\qquad \xi_{ij} = i\Delta\xi,\quad\eta_{ij} = j\Delta\eta,
     \label{eq:mt_discrete_map}
\end{equation}
holds. In all the simulations we used~$\Delta\xi = \Delta\eta = 1$.
The NN $\hat{u}$ receives the grid points $(\xi_{i,j}, \eta_{i,j})\in\Sigma_h$ as inputs and predicts the primitive variables
\begin{equation}
    \begin{aligned}
    &\hat{\bm{u}}: \Sigma \rightarrow \Omega\\
    &\hat{\bm{u}}(\xi_{ij},\eta_{ij}) = (\rho(\xi_{ij},\eta_{ij}), u(\xi_{ij},\eta_{ij}), v(\xi_{ij},\eta_{ij}), p(\xi_{ij},\eta_{ij})).
    \end{aligned}
\label{eq:prediction_nn}
\end{equation}
Predictions in $\Omega_h$ are obtained, using $\hat{\bm{u}}\left ( \mathcal{F}(x_{ij},y_{ij})\right )$, with $(x_{ij},y_{ij})\in \Omega_h$. Contrary to classical PINNs we do not obtain predictions at arbitrary coordinates. In general, this is not critical. For example, interpolation can be used to obtain predictions at arbitrary points $(x, y)\in\Omega$. In practice, linear interpolation is often sufficient.\par
The calculation of the loss function (see Sec.~\ref{sec:methods_loss_function}) requires derivatives of the primitive variables~\eqref{eq:prediction_nn} with respect to~$(x, y)$. The mesh transformation is, however, only defined in terms of the discrete mapping of Eq.~\eqref{eq:mt_discrete_map} and we have no analytical expression at hand. Therefore, we cannot directly use automatic differentiation to calculate the derivatives. Instead,  using the differentiability of~$\mathcal{F}$, the derivatives can be expressed in terms of the derivatives in $\Sigma$. The relations for the reconstruction of the first and second order derivatives are derived in~\cite{Anderson.1995, MatthewMcCoy.1980} and listed in~\ref{sec:derivatives_in_mesh_trafo}. For this reconstruction, so-called inverse metrics are required. These are the derivatives of the physical coordinates $(x, y)$ with respect to the curvilinear coordinates $(\xi, \eta)$ and can be approximated in $\Sigma_h$ using finite differences.

\subsection{Loss Function}
\label{sec:methods_loss_function}
To solve a steady-state boundary-value problem governed by the Euler equations using a PINN, we consider a loss function
\begin{equation}
    \mathcal{L} =\lambda_\mathrm{res} \mathcal{L}_\mathrm{res} + \mathcal{L}_\mathrm{bdr}.
    \label{eq:loss_function}
\end{equation}
The scalar weighting factor $\lambda_\mathrm{res}>0$ can be used to scale the magnitude of the residual loss $\mathcal{L}_\mathrm{res}$ and the boundary loss $\mathcal{L}_\mathrm{bdr}$, possibly avoiding imbalances of the losses during training~(see for example~\cite{Wang.2021, Jin.2021, Maddu.2022, Wang.2023}). Adaptive weighting algorithms may be used to adapt~$\lambda_\mathrm{res}$ automatically. For the problems considered in this article we did not observe a benefit using such algorithms compared to a constant value of $\lambda_\mathrm{res}>0$.\par

To complete the boundary-value problem we impose farfield boundary conditions at the outer part of the domain~(blue line in Fig.~\ref{fig:computational_vs_physical_domain}). Using nondimensionalized
variables~(see for example~\cite[ch. 2.3]{Langer.2018}) these are given by:
\begin{equation}
    \begin{aligned}
        \rho_{\infty} & = 1,\\
        (\rho u)_\infty & = \sqrt{\kappa} M_\infty \cos(\alpha),\\
        (\rho v)_\infty & =  \sqrt{\kappa}M_\infty\sin(\alpha),\\
        (\rho E)_{\infty} & = \frac{1}{\kappa - 1}  + \frac{\kappa M_{\infty}^2}{2}.
    \end{aligned}
    \label{eq:farfield_bc}
\end{equation}
Here~$M_{\infty}$ is the inflow Mach number and~$\alpha$ the angle of attack. From Eq.~\eqref{eq:farfield_bc}
follows the vector of primitive variables in the far field $\bm{w}_\infty = (\rho_\infty, u_\infty, v_\infty, p_\infty)$ follows. On the boundary of the airfoil a slip-wall boundary condition is imposed
\begin{equation}
    \bm{q}(\xi, \eta)\cdot\bm{n}(\xi, \eta) = 0, \qquad (\xi, \eta)\in\partial\Sigma_\mathrm{af}, 
\end{equation}
where $\bm{n}(\xi, \eta)$ is the outward facing normal vector of the airfoil, $\partial\Sigma_\mathrm{af}$ is the lower boundary of the computational domain and~$\bm{q}=(u, v)$ is the velocity vector. The computational domain features a left boundary $\partial\Sigma_\mathrm{l}$ and right boundary $\partial\Sigma_\mathrm{r}$, for which we have to impose periodic boundary conditions, since it models a physical domain which is wrapped around the airfoil
\begin{equation}
    \begin{aligned}
        \hat{\bm{u}}(\xi_\mathrm{l}, \eta) &\equiv \hat{\bm{u}}(\xi_\mathrm{r}, \eta),\\
        (\xi_\mathrm{l}, \eta)\in \partial\Sigma_\mathrm{l}&,\;\;\;(\xi_\mathrm{r}, \eta)\in \partial\Sigma_r.
    \end{aligned}
     \label{eq:periodic_bc}
\end{equation}
 Note that this is not a physical boundary conditions of the Cauchy problem in the classical sense and is only to added to precipitate continuity between the left and right boundary in the computational domain.
Altogether, Eqs.~\eqref{eq:farfield_bc}-\eqref{eq:periodic_bc} describe the loss term~$\mathcal{L}_\mathrm{bdr}$ in~\eqref{eq:loss_function},
\[
\mathcal{L}_\mathrm{bdr} = \mathcal{L}_\infty + \mathcal{L}_\mathrm{af} + \mathcal{L}_\mathrm{lr},
\]
evaluated at the respective boundaries of the computational grid $\Sigma_h$:
\begin{equation}
    \begin{aligned}
        \mathcal{L}_\infty & = \frac{1}{N_\xi} \sum_{i=1}^{N_{\xi}} (\hat{\bm{u}}(\xi_{i, N_\eta}, \eta_{i, N_\eta}) - \bm{w}_\infty)^2,\\ \mathcal{L}_\mathrm{af} & = \frac{1}{N_\xi} \sum_{i=1}^{N_{\xi}}(\bm{q}(\xi_{i, 1}, \eta_{i, 1}) \cdot \bm{n}(\xi_{i, 1}, \eta_{i, 1}) )^2, \\ 
        \mathcal{L}_\mathrm{lr} & = \frac{1}{N_\eta} \sum_{j=1}^{N_{\eta}}(\hat{\bm{u}}(\xi_{1, j}, \eta_{1, j}) - \hat{\bm{u}}(\xi_{N_\xi, j}, \eta_{N_\xi,j}))^2.
    \end{aligned}
\end{equation}
For the residual loss~$\mathcal{L}_\mathrm{res}$, we use the squared, left hand side of the differential equation, and, as described in Sec.~\ref{sec:methods_overview}. The loss is evaluated for all $N=N_{\xi}N_{\eta}$ grid points $\in\Omega$,
\begin{equation}
    \begin{aligned}
        \mathcal{L}_\mathrm{res} =\frac{1}{N} \sum\limits_i \bigg[& \frac{\partial\bm{F}_x}{\partial x} (\xi_i, \eta_i) + \frac{\partial\bm{F}_y}{\partial y} (\xi_i, \eta_i)+ \bm{D}(\bm{W}(\xi_i, \eta_i), \xi_i, \eta_i) \bigg] ^2\, ,
    \end{aligned}
\label{eq:loss_residual}
\end{equation}
where the scalar functions~$\mu$ represents the AV in x- and y-direction. As mentioned in Sec.~\ref{sec:methods_mesh_trafo}, the required derivatives are computed as described in~\ref{sec:derivatives_in_mesh_trafo}. Note that an additional viscous term $\bm{D}(\bm{W})$ is added to the PDE.\par
\subsection{Artificial Viscosity}
\label{sec:methods_artificial_viscosity}
The dissipative term in Eq.~\eqref{eq:loss_residual} is defined as:
\begin{equation}
    \bm{D}(\bm{W}(\xi, \eta), \xi, \eta)=\left(\mu_x(\xi, \eta)\frac{\partial^2}{\partial x^2} + \mu_y(\xi, \eta)\frac{\partial^2}{\partial y^2}\right)\bm{W}(\xi, \eta). 
    \label{eq:dissipative_term}
\end{equation}
It is required since PINNs use a NN as the ansatz function for the solution. For standard activation functions, NNs are continuous. Chaumet~and~Giesselmann~\cite{Chaumet.2024} recently showed for a continuous approximation $\hat{u}$ of a shock solution $u$ that the $L_2$-norm based residual loss (such as Eq.~\eqref{eq:loss_residual}) used in PINNs does not reduce when the approximation of the shock is improved. More specifically, for a Riemann problem in $\Omega\times(0, T)$ for the Burgers' equation and with the residual operator $\mathcal{R}(\cdot)$, the following inequality can be derived~\cite{Chaumet.2024}:
\begin{equation}
		\|\mathcal{R}(\hat{u}_\epsilon)\|_{L^2(\Omega\times(0, T))} \geq \dfrac{\sqrt{T}(1 +\epsilon)}{\sqrt{2\epsilon}},
        \label{eq:residual_inequality_loss}
\end{equation}
where $\epsilon\geq0$ is a small scalar and $\hat{u}_\epsilon\rightarrow u$ for $\epsilon\rightarrow 0$. According to Eq.~\eqref{eq:residual_inequality_loss}, the residual loss must at least increase with $1/\sqrt{\epsilon}$ as $\epsilon\rightarrow 0$. Hence, the commonly used residual loss is not suitable for obtaining shock solutions, without additional modifications.\par
Here we circumvent this issue via the addition of~\eqref{eq:dissipative_term} in the vicinity of shocks, resulting in a smoothed approximation of the shock. Hence, the standard $L_2$ based loss can be used.\par
The accuracy and stability of the approach depend significantly on the construction of Eq.~\eqref{eq:dissipative_term}. On the one hand,~$\mu$ needs to be sufficiently large to stabilize the training, in particular near discontinuities. On the other hand~$\mu$ will smooth out shocks and potentially obstruct convergence when it is too large.\par 
Inspired by~\cite{JAMESON.1981}, we define the \textbf{scalar~AV}
\begin{equation}
    \mu_x(\xi, \eta) = \mu_y(\xi, \eta) = \nu\;s(\xi, \eta) \;(c(\xi, \eta) + \|\bm{q}(\xi, \eta)\|).
    \label{eq:scalar_viscosity}
\end{equation}
Here,~$c + \|\bm{q}\|$ is the spectral radius of the flux jacobians $A = \partial F_x/\partial W$ and $B=\partial F_y /\partial W$~\cite{Langer.2018}, and it is weighted by a global scalar factor $\nu$ multiplied with a sensor function~$s$ constructed in Sec.~\ref{sec:methods_sensor}. The sensor is limited to the range $s(\xi, \eta)\in[0, 1)$ and identifies the regions of the domain which requires AV.\par
Eq.~\eqref{eq:scalar_viscosity}, applies the same viscosity magnitude to all four equations. However, in general the momentum variables $\rho u$ and $\rho v$ are continuous across shocks. Hence, the momentum equations do not need to be smoothed for the PINN to approximate $\rho u$ and $\rho v$ exactly. Consequently, the AV can be applied to the density and energy equations only, reducing dissipativeness in the momentum equations. We define the \textbf{density~and~energy~AV}
\begin{equation}
\mu_x(\xi, \eta) = \mu_y(\xi, \eta) = \nu\;s(\xi, \eta)\;I_{\rho, \rho E}\;(c(\xi, \eta) + \|\bm{q}(\xi, \eta)\|),
\label{eq:rho_rhoE_viscosity}
\end{equation}
with $I_{\rho, \rho E} = \mathrm{diag}(1, 0, 0, 1)$.
To further reduce dissipation, the propagation direction of waves needs to be taken into account. In classical CFD methods, this is achieved with so called matrix valued AV~\cite{Turkel.1989, Swanson.1992}, where the spectral radii in the dissipative terms are replaced with matrices. For PINNs, we define the \textbf{matrix~valued~AV}:
\begin{equation}
    \begin{aligned}
                \mu_x(\xi, \eta) &= \nu\;s(\xi, \eta)\;I_{\rho, \rho E} |A| \\
                \mu_y(\xi, \eta) &= \nu\;s(\xi, \eta)\;I_{\rho, \rho E} |B|,
    \end{aligned}
    \label{eq:matrix_viscosity}
\end{equation}
with $|A|= T |\Lambda_x| T^{-1}$ when $A = T\Lambda_x T^{-1}$ and $|B|$ accordingly for the y-direction. The matrix $\Lambda_x$ essentially represents the diagonal matrix of the eigenvalues of $A$. However, the eigenvalues are limited to avoid instabilities at sonic-lines and stagnation points, as shown in~\cite{Swanson.1992}. Further details are summarized in~\ref{sec:apx_matrix_diss}. 
\subsection{Sensor Function for normal shock waves}
\label{sec:methods_sensor}
\begin{figure}[tb]
    \centering
    \includegraphics[width=\linewidth]{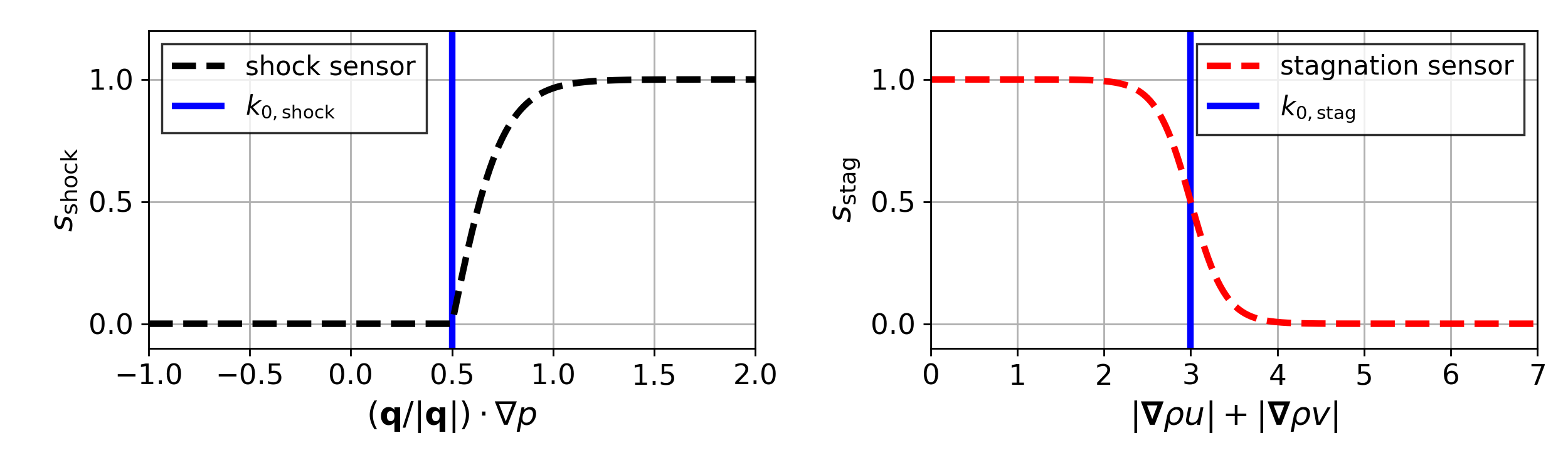}
    \caption{Schematic illustration of $s_\mathrm{shock}$ (left) for $k_{\mathrm{shock}, 1}=4,\; k_{\mathrm{shock}, 0} = 0.5$ and $s_\mathrm{stag}$ (right) for $k_{\mathrm{stag}, 1}=5,\;k_{\mathrm{stag}, 0}=3$.}
    \label{fig:sensors_schematic}
\end{figure}

Across normal shocks, an abrupt change in the flow field in flow direction is observed. The NN is a continuous ansatz function. It approximates a discontinuity by a continuous possibly very sharp transition window with high gradients of the flow variables. In general, the pressure over normal shock waves increases in the flow direction. Hence, the dot product of the normalized flow velocity $\bm{q}/\|\bm{q}\|$ and the pressure gradient $\nabla p$, will be positive in the region of a compression wave. This is exploited to construct the shock sensor
{\small
    \begin{equation}
        s_\mathrm{shock}(\xi, \eta) = \tanh\left( \max \left(0, k_{\mathrm{shock}}^{(1)} \left(\frac{\bm{q}(x,y)}{|\bm{q}|} \cdot  (\bm{\nabla}p(\xi, \eta)) - k_{\mathrm{shock}}^{(0)}\right)\right)\right).
        \label{eq:sensor_pressure_grad}
    \end{equation}}
Note that the derivatives are taken for the cartesian coordinates in $\Omega$ ($\bm{\nabla} = (\partial/\partial x , \partial /\partial y)$) and calculated as shown in~\eqref{eq:apx_mesh_metrics}. The hyperbolic tangent and the inner max-function limit the sensor to $s(\xi, \eta)_\mathrm{shock}\in (1, 0] \text{ for } (\xi, \eta)\in\Sigma$. The second parameter $k_{\mathrm{shock}, 1}$ controls the steepness of $s(x, y)$ for increasing gradients. Overall, in regions of {high positive pressure gradients in flow direction} the sensor becomes $s\approx 1$. Everywhere else, the sensor is turned off ($s=0$). \\
While the sensor is designed to identify normal shock waves, strong pressure gradients in flow direction also occur at stagnation points, for example at the leading edge of the airfoil. 
Our experiments show that convergence is obstructed when the sensor is active at stagnation points. Therefore, we extend the sensor by a second factor which removes the stagnation points. Here, we use the fact that the momentum in x- and y-direction is conserved across normal shock waves. Hence, we can assume that the gradients of $\rho u$ and $\rho v$ are small in the region of the shock. The momentum will however experience strong gradients near stagnation points. We therefore propose the stagnation point sensor 
\begin{equation}
    s_\mathrm{stag}(\xi, \eta) = \mathrm{sig}(k_{\mathrm{stag}}^{(1)}(k_{\mathrm{stag}}^{(0)} - (|\bm{\nabla}\rho u| + |\bm{\nabla}\rho v|)). 
    \label{eq:sensor_stagnation}
\end{equation}
Here, the sigmoid function $\mathrm{sig} (x) = 1/(1+\exp(-x))$ ensures that $s_\mathrm{stag}(\xi, \eta)\in(0, 1)$. Again, the parameters $k_{\mathrm{stag}, 1}$ and $k_{\mathrm{stag}, 0}$ can be used to adapt the steepness of the switch and the threshold. Overall, near the stagnation point, the sensor becomes $s_\mathrm{stag}\approx 0$ and everywhere else it will be $s_\mathrm{stag}\approx 1$. The complete sensor function $s(\xi, \eta)$ is the product of the shock and the stagnation-point sensor
\begin{equation}
    s(\xi,\eta) = s_\mathrm{shock}(\xi, \eta)s_\mathrm{stag}(\xi, \eta).
    \label{eq:sensor_combined}
\end{equation}
Fig.~\ref{fig:sensors_schematic} illustrates the two sensor components. In Sec.~\ref{sec:results_baseline} we also show $s(x, y)$ in a practical example.
The recommended values for $k^{(0)}_{\mathrm{shock}}, k^{(1)}_{\mathrm{shock}}, k^{(0)}_{\mathrm{stag}}$ and $k^{(1)}_{\mathrm{shock}}$ are shown in~\ref{sec:appendix_hyperparameters} and are kept constant for all experiments. To calculate the gradients in Eqs.~\eqref{eq:sensor_pressure_grad}-\eqref{eq:sensor_stagnation}, we also need the derivatives with respect to $x$ an $y$. How to obtain these derivatives using the inverse mesh metrics is described in Sec.~\ref{sec:methods_mesh_trafo}.

\subsection{Trainable layers}
\label{sec:methods_trainable_layers}
The presented PINN model uses a fully connected NN~$ \hat{\bm{u}}:\mathbb{R}^{N_0}\to \mathbb{R}^{N_D}$ with a depth of $D$ layers. Each layer $k = 1\dots D$ is defined by a function $f^k$ with 
\begin{equation}
    \begin{aligned}
        \hat{\bm{u}} &= f^D\circ \dots  \circ f^k \circ \dots \circ f^1(\bm{r}),\\
        f^k(\bm{r}) &= \sigma^k (w^k \bm{r} - b^k).
    \end{aligned}
\end{equation}
The trainable parameters of the model are the bias vector $b^k\in \mathbb{R}^{N_k}$ and the weight matrices $w\in\mathbb{R}^{N_k}\times\mathbb{R}^{N_{k-1}}$. As the activation function, the presented models uses the 
the layer-wise adaptive activation functions with a hyperbolic tangent, as introduced by Jagtap~et al.~\cite{Jagtap.2020b}
\begin{equation}
    \begin{aligned}
        \sigma^k(x)&=\tanh(n\omega_\sigma^k\cdot x),\qquad k=1,2\ldots,  D-1,
    \end{aligned}
\end{equation} 
where $n$ is a constant scaling factor and $\omega_\sigma^k$ is an additional trainable parameter. The computational effort per epoch is only marginally increased because only one parameter per layer is added.\par
Besides the adaptive activation functions, the weight normalization methodology is also used for the model. Weight normalization is consistently able to outperform the classical fully connected architecture~\cite{Salimans.2016}. The idea is to replace each vector in the weighting matrix of each layer $w^k$ by a normalized 
vector~$\bm{z}/\|\bm{z}\|$ and a scalar magnitude $g$, that is
\begin{equation}
    w^k = (\bm{w}_1, \bm{w}_2, \dots, \bm{w}_i,\dots, \bm{w}_{N_{k-1}})^T,\quad \bm{w}_i \longrightarrow g_i \frac{\bm{z}_i}{\|\bm{z}_i\|}.
\end{equation}
This often improves the conditioning of the optimization problem. Hence, it has shown to be useful for PINNs where convergence is often slow. The factorization adds a number of additional parameters $g$ on the order of the number of neurons. Therefore, the increase in computational effort per epoch is again comparatively small.  
\subsection{Fourier-Feature Embedding}
\label{sec:methods_fourier_features}
The trainability of deep NNs generally suffers from a phenomenon called \textit{spectral bias}. During the training of a NN, the low frequencies of the target function are generally learned first. The higher frequency modes are only learned later during training~\cite{Rahaman.2018, Xu.2022}. In practice, this means that the convergence on multi-scale problems is generally poor and that they are hard to solve for PINNs. The random Fourier-feature embedding~\cite{Tancik.2020, Wang.2021b} aims to overcome these limitations. An additional non-trainable layer is introduced which encodes the input vector of the NN using a set of sinusoidal functions with randomly sampled frequencies:
\begin{equation}
    \begin{aligned}    
        &\bm{\hat{u}}_\mathrm{fe}(\bm{r}): \mathbb{R}^{D_0} \longrightarrow \mathbb{R}^{2D_\mathrm{fe}},\\
        &
        \begin{pmatrix}
        \hat{u}_{\mathrm{fe}, i}\\
        \hat{u}_{\mathrm{fe}, i+1}\\
        \vdots
        \end{pmatrix}= 
        \begin{pmatrix}
            \sum_{j=1}^{D_0} \sin(\phi_{i, j}r_j)\\
            \sum_{j=1}^{D_0} \cos(\phi_{i, j}r_j)\\
            \vdots
        \end{pmatrix},\quad
        &i = 2k - 1,\;\;k = 1\dots D_\mathrm{fe}.
    \end{aligned}
\end{equation}
The frequencies $\phi_{i, j}$ are typically sampled from a normal distribution $\mathcal{N}(0, \sigma^2_\mathrm{\phi})$. The standard deviation $\sigma_\phi$ of the distribution is a hyperparameter of the resulting model. Empirical tests on the models in this work showed good performance for $\sigma_\phi\in [1, 3)$. For all presented results we used $\sigma_\phi=1$.  
\subsection{Implementation Details}
The presented PINN framework uses the \textit{Surrogate Modeling for Aero Data Toolbox in Python} (SMARTy)~\cite{Bekemeyer.2022} for the implementation of the NN. In this case, we use the PyTorch~\cite{Paszke.03.12.2019} backend of SMARTy, which provides crucial parts of the model and the training algorithms, as well as a straightforward calculation of the derivatives with automatic differentiation. Furthermore, we use the PyTorch implementation of the low memory Broyden-Fletcher-Goldfarb–Shanno algorithm (L-BFGS), with a maximum number of 1000 inner iterations. We perform mini-batch training. However, since the PyTorch implementation of L-BFGS is not natively suited for stochastic loss functions, we reset the optimizer after each outer iteration to avoid complications with the approximation of the Hessian as the underlying data changes for each batch. Note that the training time can vary between runs, since the inner iterations are stopped when the convergence criteria are reached. A summary of all optimizer parameters is shown in Tab.~\ref{tab:training_parameters_baseline}.\par
\begin{figure}[tb]
    \centering
    \includegraphics[width=0.8\linewidth]{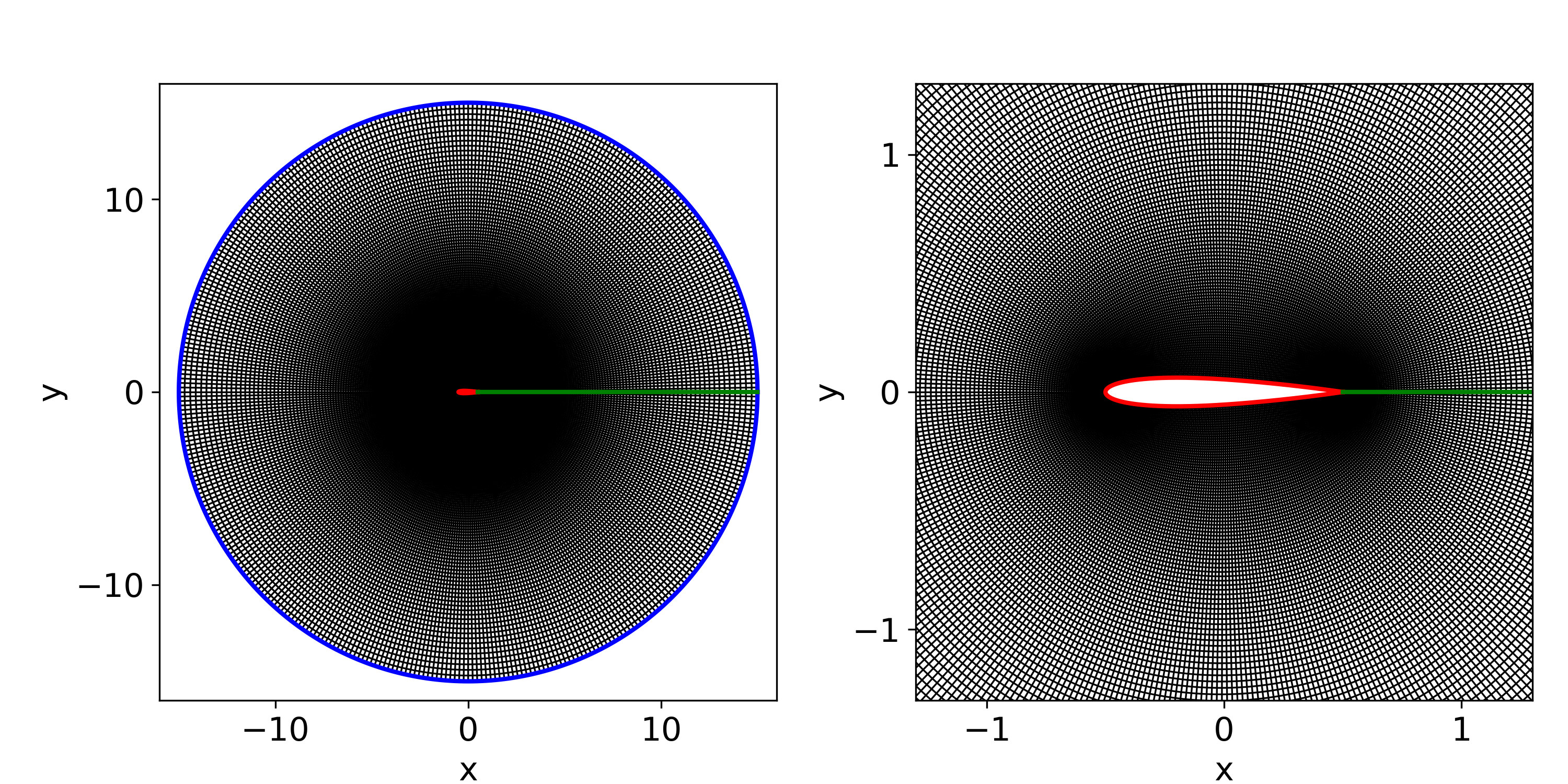}
    \caption{The used curvilinear grid. The left image shows the entire grid. The right image shows a close-up of the airfoil. The boundaries of the computational grid are color coded, similarly to~Fig.~\ref{fig:computational_vs_physical_domain}}
    \label{fig:grid}
\end{figure}

The implementation of the mesh transformation is heavily inspired by the code provided by Cao~et~al.~\cite{Cao.2024b}. Similarly to them, the elliptical grids are generated, using a finite difference solver for the grid point locations as described in \cite[pp. 192-200]{Anderson.1995}. For the PINN, we use an O-type grid with $400\times200$ points. The grid is shown in Fig.~\ref{fig:grid}. Compared to~\cite{Cao.2024b}, where a $200\times100$ point grid is used, we have seen that the shock is more stable and that the method performs more consistently, when a higher resolution grid is used. The inverse metric terms (Eqs.~\eqref{eq:apx_mesh_metrics}) are calculated using higher order finite differences~\cite{Fornberg.1988}. The second order mesh metrics were calculated based on the relations derived in~\cite{MatthewMcCoy.1980}.\par
The calculation of the second order derivatives is significantly more costly than for the first order derivatives. The second order derivatives are however only required in regions where viscosity is applied i.e. where $s>0$. Therefore, we first calculate all first order derivatives required for the loss function for all grid points. This includes the pressure gradients which are required to calculate the sensor function. Once the sensor function (Eq.~\eqref{eq:sensor_combined}) is calculated, we check at which grid points the sensor exceeds a threshold of $10^{-4}$. We only calculate the viscous terms for these points and set them to $0$ for all other grid points, reducing the overall computational cost per training iteration.\par Note also that, the sensor function $s(x, y)$ is detached from the computational graph. This is due to the fact that the PINN should learn the solution to the inviscid PDEs and not minimize for the best sensor position to lower the loss function. The sensor position should be solely determined by the location of the shock according to the PDEs themselves.\par
For the calculation of the loss function, we also employ the weighting of the residuals based on the cell volume, as proposed by Song~et~al.~\cite{Song.2024}. To do that, we can modify Eq.~\eqref{eq:loss_residual}, and multiply each term in the sum with a weighting factor, which increases the weight of bigger cells on the loss to avoid a bias towards high point density areas in the domain.\par
The combination of the mesh transformation with the volume weighting leads to a residual loss term which is typically orders of magnitude lower than the boundary loss term. Therefore, we apply a large constant scaling of $\lambda_\mathrm{res}=2\cdot10^4$ (see Eq.~\eqref{eq:loss_function}). A summary of the used hyperparameters of the models is shown in Apx.~\ref{sec:appendix_hyperparameters} 
\label{sec:methods_implementation_details}
\subsection{PINN Error Evaluation}
\label{sec:methods_error_metrics}
Among other quantities, we we consider the pressure coefficient $C_p$ and the local Mach number $M$,
\begin{equation}
    C_p = \frac{p - p_\infty}{\frac{1}{2}\rho_\infty \|\bm{q}_\infty\|^2},\quad M=\frac{\|\bm{q}\|}{c}\, .
\end{equation}
for the evaluation of the results. Reference values are obtained, using a finite volume solver (see Sec.~\ref{sec:reference_simulation_coda}) 
Three different error metrics are evaluated for a quantitative comparison. We consider the mean absolute error
\begin{equation}
	MAE = \frac{1}{n} \sum_{i=0}^{n} |\hat{u}(x_i, t_i) - u(x_i, t_i)|,
\end{equation}
the relative mean absolute error
\begin{equation}
	RMAE = \frac{MAE}{\max_i (u_i) - \min_i(u_i)},
\end{equation}
and the coefficient of determination (or $R_2$ score)
\begin{equation}
	R_2 = 1 - \frac{(\hat{u}(x_i, t_i) - u(x_i, t_i))^2}{(\langle u(x_i, t_i)\rangle - u(x_i, t_i))^2},
\end{equation}
with reference values $u_i$ and predicted values $\hat{u}_i$. The relative $L_2$ error is not evaluated here, as it is not well suited as a relative metric for quantities such as $C_p$ with values centered around $C_p\approx0$. The operator $\langle \cdot \rangle$ indicates the mean of all $i$ points.

\section{Results - Artificial Viscosity Validation}
\label{sec:results_baseline}
Reference results are obtained, using a finite volume solver (see Sec.~\ref{sec:reference_simulation_coda}). We consider the NACA~0012 airfoil, as the geometry. 
Four different test cases, with different far-field Mach numbers $M_\infty$ and angles of attack $\alpha$ are considered, as shown in Tab.~\ref{tab:test-cases}.
\begin{table}[tb]
    \caption{Parameter configuration for different transonic test cases.}
    \label{tab:test-cases}
    \centering
    \begin{tabular}{|c|c|c|c|c|}
        \hline
         &  case A & case B & case C & case D\\
        \hline
        Mach number $M_\infty$ & 0.7 & 0.72 & 0.75 & 0.78 \\
        angle of attack $\alpha\;[^\circ]$  & 4 & 3  & 2 & 3 \\
        \hline
    \end{tabular}
\end{table}
They represent typical combinations of $M_\infty$ and $\alpha$, encountered on commercial aircraft in the transonic regime and feature a distinct normal shock wave on the upper side of the airfoil. Considering that the sweep angle of aircraft wings reduces the effective Mach number present at a local airfoil section, these cases emulate typical effective Mach numbers and angles of attack encountered on turbo-fan aircraft. Test case~D represents an edge case scenario at a comparatively high Mach number. For a sweep angle of $30^\circ$ it corresponds to a flight Mach number of $M_\infty\approx0.9$ and a high angle of attack for such speeds.\par
The failure of PINNs to predict shocks, without additional modifications, has been rigorously confirmed in numerous experiments~\cite{Fuks.2020, Patel.2022, Coutinho.2023, Wagenaar.2023, Cao.2024b, Jin.2024} and was recently also confirmed theoretically~\cite{Chaumet.2024}. In addition, different variants of AV have been used as an effective countermeasure in various works~\cite{Fuks.2020, Patel.2022, Wassing.2024, Wagenaar.2023}. In Apx.~\ref{sec:apx_pinn_without_av}, we confirm again that without AV, our model also fails to predict normal shocks, encountered in transonic flows, while in this section we showcase that this issue can be overcome using AV.\par
For all experiments, we use a neural network with Fourier embedding, weight normalization and adaptive activations~(see Sec.~\ref{sec:methods_trainable_layers}). A total of five models with different AV variants are analyzed and as a sanity check, we also run simulations without AV (i.e. $\nu=0$). The model variants are summarized in Tab.~\ref{tab:models_and_nus}. For the models where the stagnation sensor or the shock sensor are not active, we fall back to global AV, simply setting $s_\mathrm{stag}=1$ or $s_\mathrm{shock}=1$, respectively. For all models with AV, we perform a grid search to obtain an optimal value for $\nu$. The used values are shown in Tab.~\ref{tab:models_and_nus} and are in a similar range of $\nu\in [1.8, 2.1]\cdot 10^{-3}$. Generally, for all $\nu$ values in this range, the accuracy of the different models does not change significantly. However, since the produced viscosity differs between the models, it is reasonable to choose an optimal values of $\nu$ to ensure a fair comparison at least in this paper. For practical use, we recommend a value of $\nu=1.9\cdot 10^{-3}$ as a sensible choice for all models.\par
\begin{figure}[tb]
    \centering
    \includegraphics[width=\linewidth]{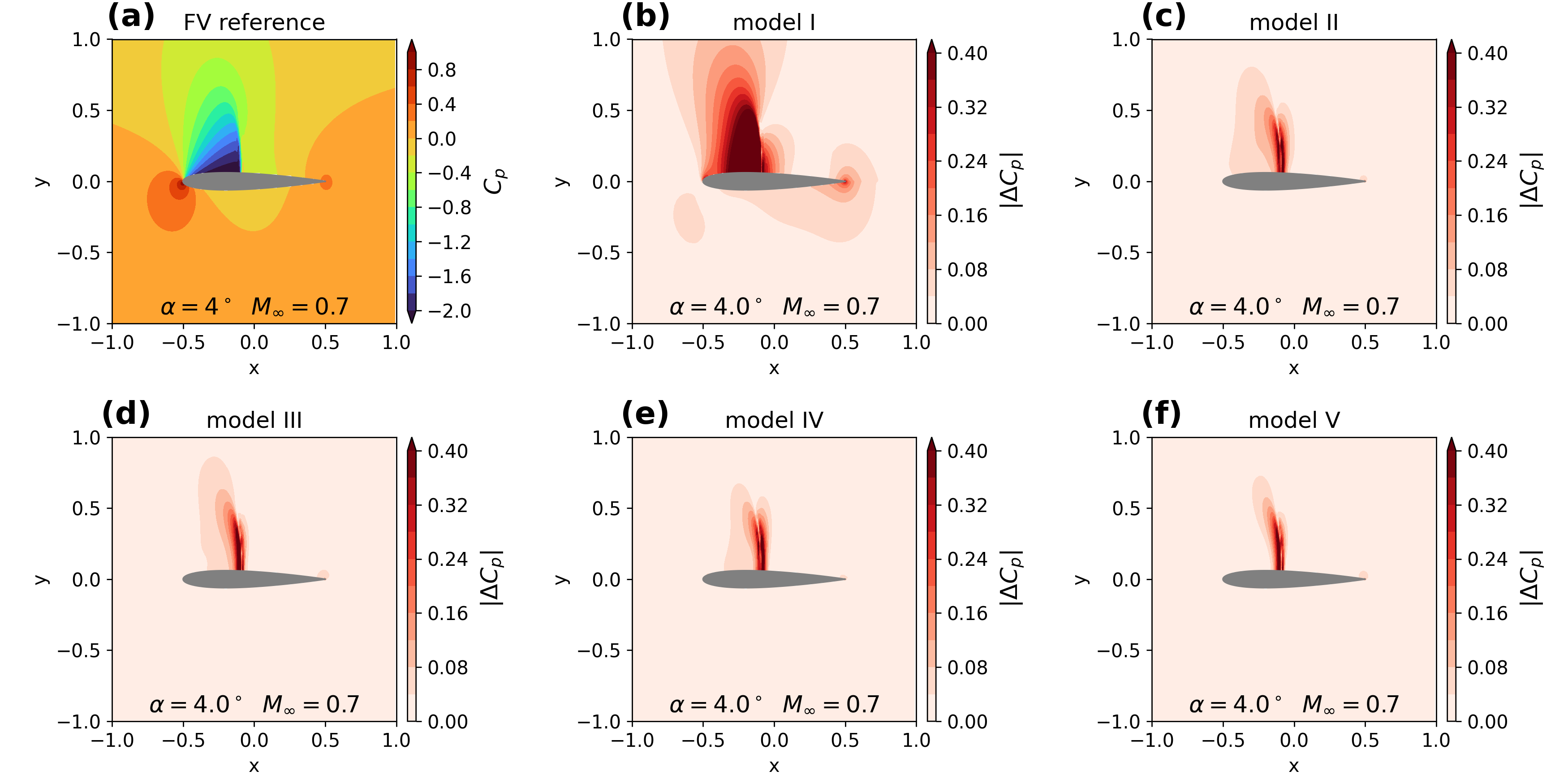}
    \caption{Predicted pressure fields in comparison to reference solution and absolute error between them for four different test cases at different Mach number $M_\infty$ and angle of attacks $\alpha$.}
    \label{fig:pressure_field_baseline}
\end{figure}
\begin{figure}[th]
    \centering
    \includegraphics[width=.8\linewidth]{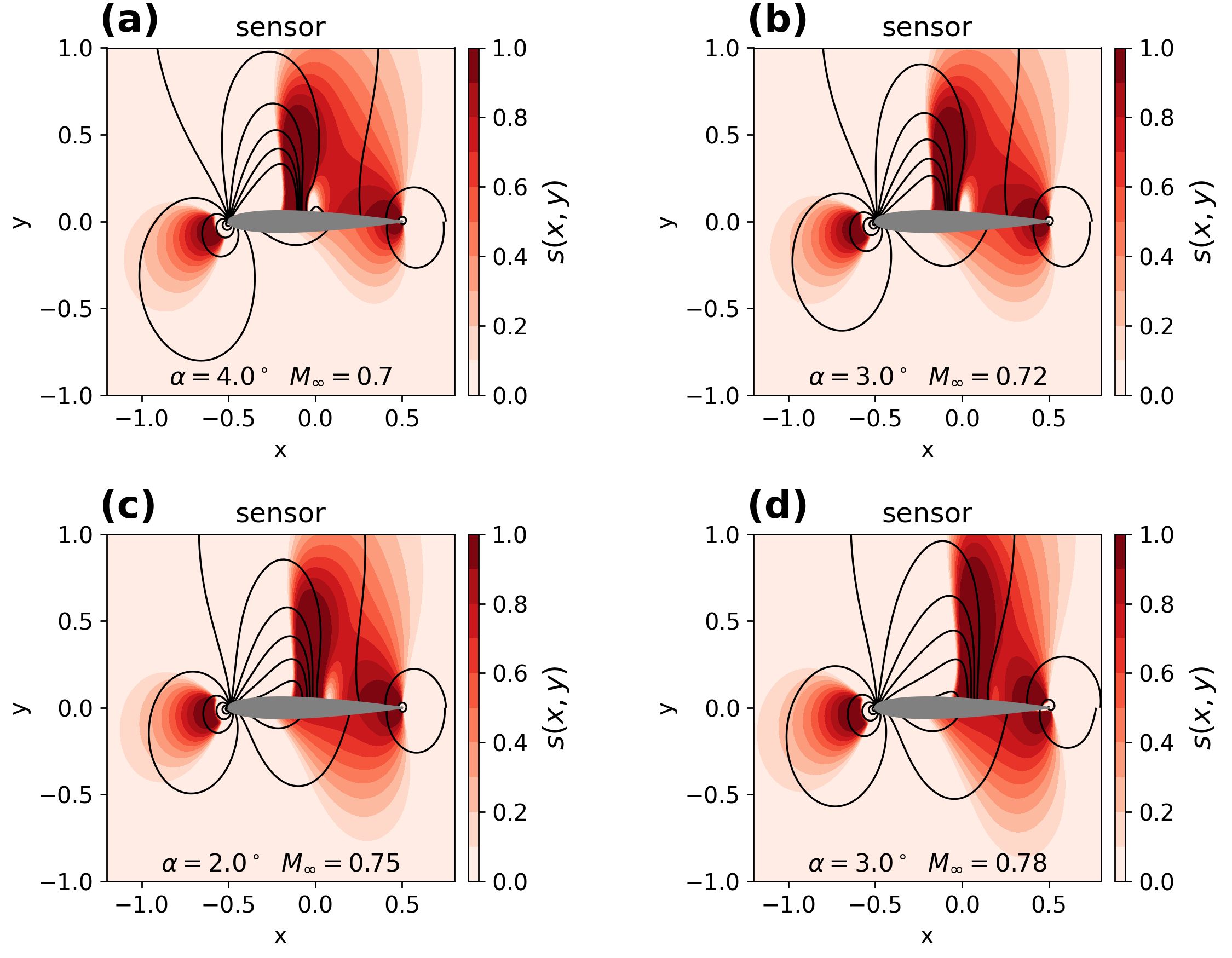}
    \caption{The sensor function $s(x, y)$ and the applied AV $\mu(x, y)$ for the three transonic test cases in Fig.~\ref{fig:pressure_field_baseline}. The contour lines show the pressure field to indicate how the sensors are affected by the pressure gradient.}
    \label{fig:sensor_field_baseline}
\end{figure}

\begin{table}[tb]
    \caption{Overview of analyzed models with different AV types.}
    \label{tab:models_and_nus}
    \centering
    \begin{tabular}{|l|c|c|c|c|c|c|}
        \hline
        model  &  $s_\mathrm{stag}$ & $s_\mathrm{shock}$ & scalar AV & $\rho$-$E$ AV & matrix AV &$\nu \cdot 10^{-3}$\\
        \hline
        O   &            &            &            &            &            & 0\\
        \hline
        I   &            &            & \checkmark &            &            & 2.1\\
        \hline
        II  & \checkmark &            & \checkmark &            &            & 1.9\\
        \hline
        III & \checkmark &            &            & \checkmark &            & 1.8\\
        \hline
        IV  & \checkmark & \checkmark &            & \checkmark &            & 2\\
        \hline
        V   & \checkmark & \checkmark &            &            & \checkmark & 1.9\\
        \hline
    \end{tabular}
\end{table}
Initially, we focus on test case~A and analyze the effectiveness of different AV schemes. The reference pressure field is shown in Fig.~\ref{fig:pressure_field_baseline}~(a). Figs.~\ref{fig:pressure_field_baseline}~(b)-(f) shows the absolute difference to the reference $|\Delta C_p|$ for the different AV variants.
Fig.~\ref{fig:pressure_field_baseline}~(b) shows the result when a global scalar AV~\eqref{eq:scalar_viscosity} is used and no sensors are active. In this case, the PINN model completely fails to approximate the shock. However, when the stagnation point sensor~\eqref{eq:sensor_stagnation} is activated, even though the shock sensor is still set to $s_{shock}=1$, the resulting model is able to approximate the shock quite accurately, as shown in Fig.~\ref{fig:sensor_field_baseline}~(c). While the shock is still smoothed out and some inaccuracies can be seen in the low pressure region in front of the shock, the improvements due to the AV are clearly visible. Switching to the density and energy AV, the accuracy can be further increased, as shown in Fig.~\ref{fig:pressure_field_baseline}~(d). Evidently, dissipation in the momentum equations is not necessarily required to stabilize the shocks, as postulated in Sec.~\ref{sec:methods_artificial_viscosity}. The removal of the dissipation hence leads to a less dissipative but similarly stable model. When the shock sensor is activated in Fig.~\ref{fig:pressure_field_baseline}~(e), the slight inaccuracies in-front of the shock can be further reduced, since the dissipation is only activate around the shock itself. The best approximation is however obtained when using matrix-valued AV, as shown in Fig.~\ref{fig:pressure_field_baseline}, where significant errors are only visible due to the smoothing of the shock.\par
\begin{figure}
    \centering
    \includegraphics[width=\linewidth]{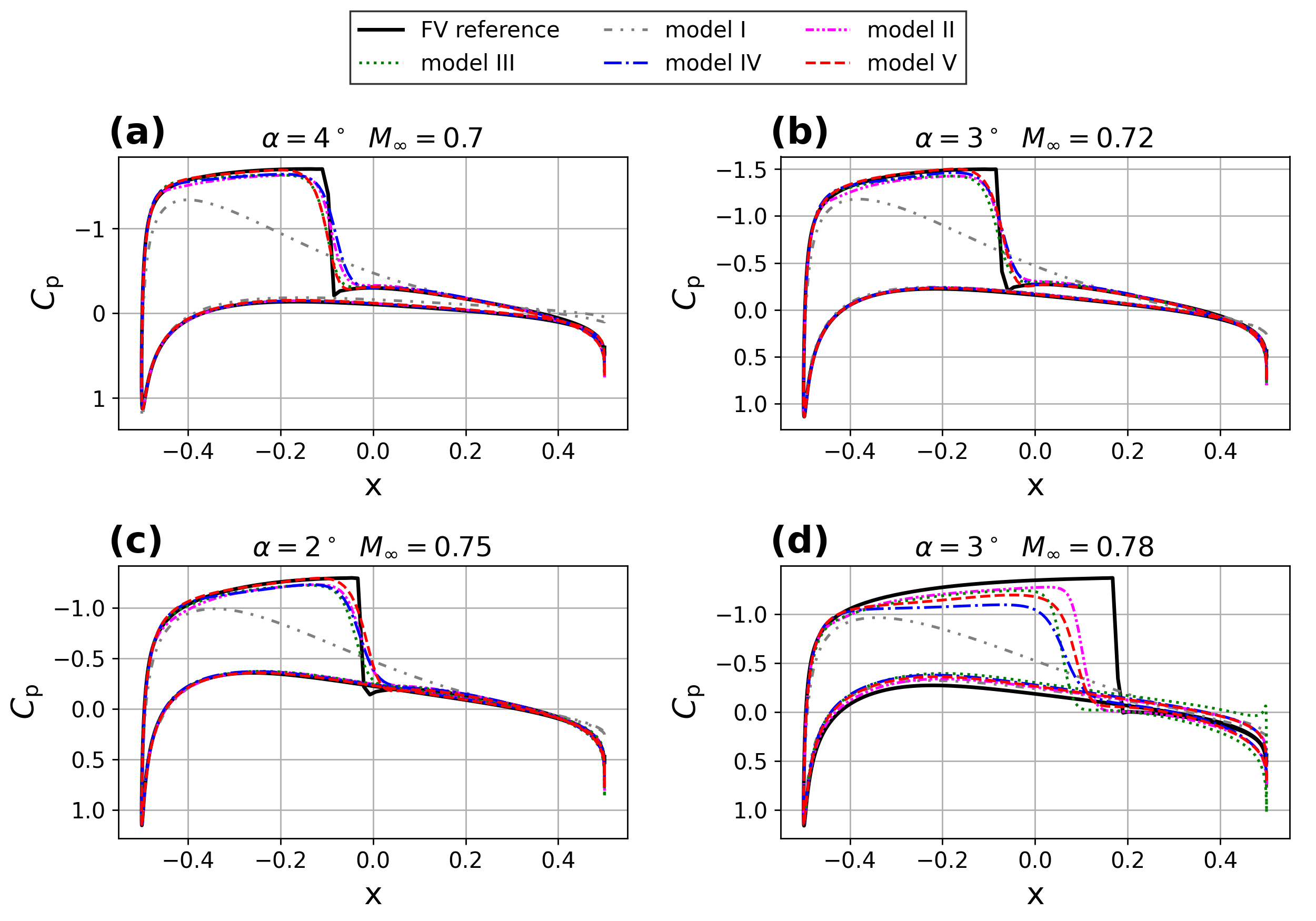}
    \caption{Surface pressure coefficient $C_p$ for case A-D. Different AV methods are compared to the finite volume reference.}
    \label{fig:surface_Cp_all_cases}
\end{figure}

To evaluate the performance of the different AV variants on the other cases, we consider surface pressure plots for each model for all four cases, as shown in Fig.~\ref{fig:surface_Cp_all_cases}.
For case B-C, we observe a similar behavior as for case~A. Model I fails to provide accurate predictions, whereas all other models perform similarly well. The matrix-valued AV used in model V performs best on case A-C.\par
However, for case~D, we see that the shock location is not accurately predicted by all AV versions and overall the predictions deviate from the reference. Model II provides the most accurate prediction, followed by model V but both model predictions are still unsatisfactory. Possible reasons for this behavior at high Mach numbers and angles of attack are discussed in Sec.~\ref{sec:conclusion_and_outlook}\par
Fig.~\ref{fig:sensor_field_baseline}~shows the sensor $s$ for model~V for case A-D. 
In addition, the contour lines of the coefficient of pressure $C_p$ are plotted. The sensor is mainly active ($s \approx 1$) near the shock, where the contour lines converge. One can also see active regions near the leading edge and close to the trailing edge. Upon close inspection, the sensor is however inactive at the stagnation points themselves due to the stagnation point sensor given in Eq.~\eqref{eq:sensor_stagnation}. The removal of the stagnation points from the sensor drastically improves the prediction accuracy, as shown in Fig.~\ref{fig:pressure_field_baseline}. In theory, the activation threshold $k^{(0)}_\mathrm{shock}$ can be increased to further limit the regions where viscosity is applied. However, empirical trials show that this does not lead to significant improvements in accuracy and slightly reduces the consistency of the method overall.\par
Finally, for a quantitative analysis of the accuracies, the $MAE$, $RMAE$, and $R_2$-score for model I-V and cases~A-D are analyzed (in comparison to the reference simulation). Each metric is calculated as the average over four different runs with different random initialization of the trainable network parameters (c.f.~\ref{sec:methods_trainable_layers}). For each run we calculate the error metrics over a box $(x, y) \in (-1, 1)\times(-1, 1)$. The points for this evaluation are uniformly distributed in the physical domain while points inside the airfoil are excluded. The error bounds are given by the standard deviation over the four training runs. Tab.~\ref{tab:errors_non_parametric_RMAE} shows the RMAE for $C_p$ and the local Mach number $M$. Italic values represent cases, where significant outliers (i.e. single runs that did not converge) negatively affected the metrics. Bold values show the best performing model. Tab.~\ref{tab:errors_non_parametric_RMAE} shows the $RMAE$, and the $MAE$ and $R_2$-score are shown in~\ref{sec:apx_error_metrics}.\par
\begin{table}[tb]
    \centering
    \caption{$RMAE$ in $[\%]$ for different test cases and field variables $C_p$ and $M$.}
    \label{tab:errors_non_parametric_RMAE}
    \begin{tabular}{|c||c|c|c|c|c|c|c|c|}
                \hline
                \rotatebox[origin=c]{90}{Model}
                & \multicolumn{2}{c|}{\makecell[c]{Case~A \\$M_\infty =0.7 $\\$\alpha = 4^\circ$}} &
                  \multicolumn{2}{c|}{\makecell[c]{Case~B \\$M_\infty =0.72 $\\$\alpha = 3^\circ$}} &
                  \multicolumn{2}{c|}{\makecell[c]{Case~C \\$M_\infty =0.75 $\\$\alpha = 2^\circ$}} &
                  \multicolumn{2}{c|}{\makecell[c]{Case~D \\$M_\infty =0.78 $\\$\alpha = 3^\circ$}}\\ \hline \hline
                0 &
                \makecell[l]{$C_p$ \\ $M$ } & 
                \makecell[c]{$4.82\pm 0.28$ \\ $3.85\pm 0.19$} &
                \makecell[l]{$C_p$ \\ $M$} & 
                \makecell[c]{$4.1\pm 0.28$ \\ $3.33\pm 0.22$} &
                \makecell[l]{$C_p$ \\ $M$} & 
                \makecell[c]{$4\pm 1$ \\ $3.1\pm 0.8$} &
                \makecell[l]{$C_p$ \\ $M$} & 
                \makecell[c]{$8.35\pm 0.11$ \\ $6.76\pm 0.11$}\\
                \hline
                I &
                \makecell[l]{$C_p$ \\ $M$} & 
                \makecell[c]{$2.2\pm 0.4$ \\ $1.79\pm 0.24$} &
                \makecell[l]{$C_p$ \\ $M$} & 
                \makecell[c]{$4\pm 2.4$ \\ $18\pm 20$} &
                \makecell[l]{$C_p$ \\ $M$} & 
                \makecell[c]{$1.79\pm 0.19$ \\ $1.59\pm 0.16$} &
                \makecell[l]{$C_p$ \\ $M$} & 
                \makecell[c]{$5.2\pm 1.9$ \\ $15\pm 22$}\\
                \hline
                II &
                \makecell[l]{$C_p$ \\ $M$} & 
                \makecell[c]{$0.48\pm 0.03$ \\ $0.474\pm 0.018$} &
                \makecell[l]{$C_p$ \\ $M$} & 
                \makecell[c]{$\mathit{6\pm 7}$ \\ $\mathit{15\pm 18}$} &
                \makecell[l]{$C_p$ \\ $M$} & 
                \makecell[c]{$1\pm 0.9$ \\ $0.9\pm 0.7$} &
                \makecell[l]{$C_p$ \\ $M$} & 
                \makecell[c]{$\mathbf{1.85\pm 0.15}$ \\ $\mathbf{1.65\pm 0.13}$}\\
                \hline
                III &
                \makecell[l]{$C_p$ \\ $M$} & 
                \makecell[c]{$\mathit{2.1\pm 2.8}$ \\ $\mathit{5\pm 8}$} &
                \makecell[l]{$C_p$ \\ $M$} & 
                \makecell[c]{$\mathit{3\pm 4}$ \\ $\mathit{10\pm 18}$} &
                \makecell[l]{$C_p$ \\ $M$} & 
                \makecell[c]{$0.46\pm 0.04 $ \\ $0.466\pm 0.029$} &
                \makecell[l]{$C_p$ \\ $M$} & 
                \makecell[c]{$5\pm $ 4\\ $13\pm 20$}\\
                \hline
                IV &
                \makecell[l]{$C_p$ \\ $M$} & 
                \makecell[c]{$0.49\pm 0.14$ \\ $0.48\pm 0.12$} &
                \makecell[l]{$C_p$ \\ $M$} & 
                \makecell[c]{$0.34\pm 0.11$ \\ $0.35\pm 0.09$} &
                \makecell[l]{$C_p$ \\ $M$} & 
                \makecell[c]{$0.56\pm 0.15$ \\ $0.54\pm 0.13$} &
                \makecell[l]{$C_p$ \\ $M$} & 
                \makecell[c]{$3.11\pm 0.14 $ \\ $2.8\pm 0.4$}\\
                \hline
                V &
                \makecell[l]{$C_p$ \\ $M$} & 
                \makecell[c]{$\mathbf{0.42\pm 0.04}$ \\ $\mathbf{0.427\pm 0.022}$} &
                \makecell[l]{$C_p$ \\ $M$} & 
                \makecell[c]{$\mathbf{0.3\pm 0.08}$ \\ $\mathbf{0.31\pm 0.05}$} &
                \makecell[l]{$C_p$ \\ $M$} & 
                \makecell[c]{$\mathbf{0.33\pm 0.11}$ \\ $\mathbf{0.35\pm 0.09}$} &
                \makecell[l]{$C_p$ \\ $M$} & 
                \makecell[c]{$3\pm 0.4$ \\ $2.7\pm 0.4$}\\
                \hline
    \end{tabular}
\end{table}
The metrics confirm that model V performs best on case A-C and that Model II performs best on case D.
Compared to the reference simulations, we generally see errors of less than one percent for all analyzed test-cases, when using model V. Moreover, it performs consistently well with a comparatively low standard deviation. Model IV performs only slightly worse. Models II-III without $s_\mathrm{shock}$ perform inconsistently, since some individual runs diverge or certain cases cannot be solved accurately at all.\par
The training time for model~V varies between $30\,\text{min}$ and $45\,\text{min}$. Due to the global viscosity, the training time of model~II is for example significantly longer, ranging from $40\,\text{min}$ to $165\,\text{min}$. The variance is caused by the convergence criteria in the inner iteration of the L-BFGS optimizer.\par
Based on the error metrics, the matrix AV, combined with the shock and stagnation-point sensor seems to be well suited for moderate Mach numbers and angles of attack in the transonic regime. In the following, the applicability of model V to a parametric setting is analyzed.
\section{Results - Parametric Model}
\label{sec:results_parametric}
\begin{figure}[p]
    \centering
    \includegraphics[width=\linewidth]{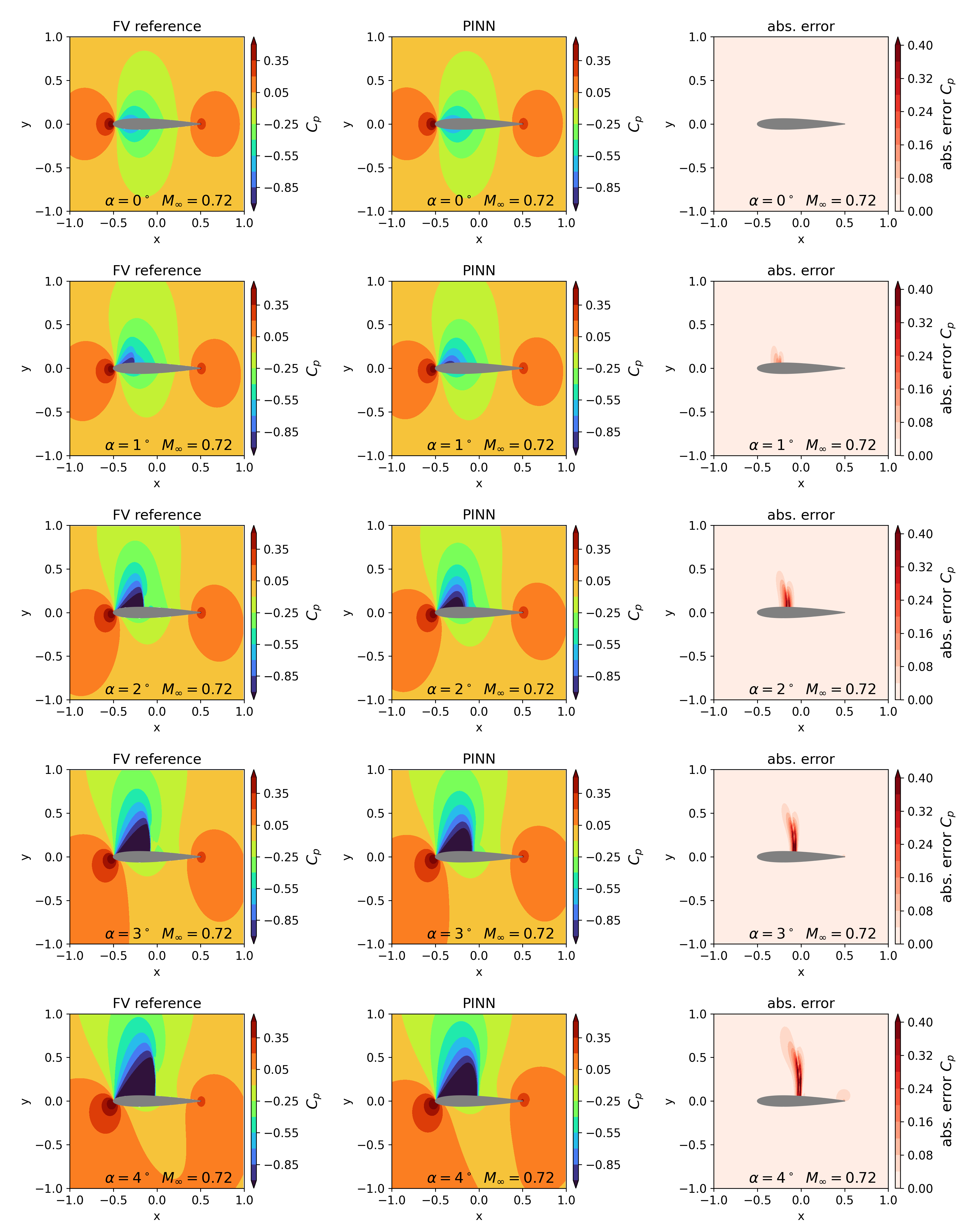}
    \caption{Predicted pressure fields by a single parametric PINN in comparison to reference solutions at different $\alpha$.}
    \label{fig:pressure_field_parameteric}
\end{figure}

The previously analyzed methodology can be adapted to parametric problems. As shown in Fig.~\ref{fig:schematic_method_overview}, parameters of the PDE can be added to the input space of the network to solve the PDE in a parametric fashion. In addition to the coordinates in the computational grid $(\xi, \eta)$, the PINN is trained in a third dimension defined by the angle of attack. The trained model can be used to make predictions for the entire training range of $\alpha$ at once. Here, we parameterize the angle of attack by randomly sampling $N_\alpha$ values for the angle of attack in the range $\alpha\in(0, 4)^\circ$. During a training iteration, each grid point is randomly combined with one of the sampled $\alpha$ values. To improve the predictions at the bounds of the parameter space, we add $N_{\alpha, \mathrm{bound}}$ additional $\alpha$ values with $\alpha=0^\circ$ and $\alpha=4^\circ$. The Mach number is kept constant at $M_\infty=0.72$. 
An overview of the resulting pressure field in comparison to the reference solution is shown in Fig.~\ref{fig:pressure_field_parameteric} for five different angles of attack. At the lowest angle of attack $\alpha\approx0$, the model approximates the fully subsonic flow accurately, even when using the AV. For $\alpha=1^\circ$, the prediction is very accurate overall, but since this shock is very weak, it not captured by the PINN. At  $\alpha=2$, the PINN can resolve the shock although it is still smoothed out. The stronger shocks at $\alpha\geq 3^\circ$ are all approximated well.\par
\begin{figure}[th]
    \centering
    \includegraphics[width=\linewidth]{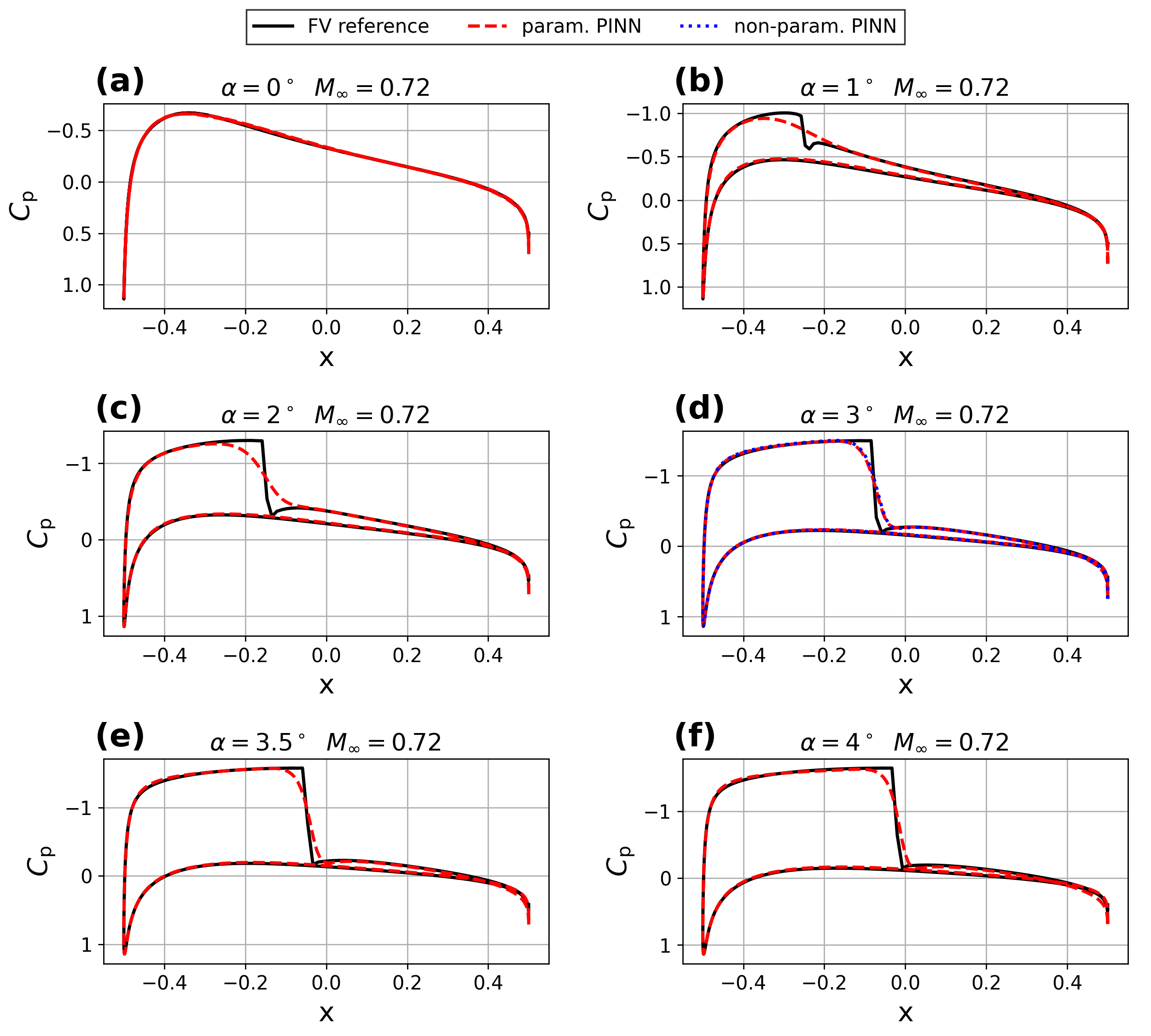}
    \caption{Predicted surface pressure distribution by a single parametric PINN in comparison to reference solutions at different $\alpha$. For $\alpha=2^\circ$ the prediction of the corresponding non-parametric PINN in Fig.~\ref{fig:pressure_field_baseline} is also shown.}
    \label{fig:pressure_profiles_parametric}
\end{figure}

Looking at the corresponding surface pressure distributions in Fig.~\ref{fig:pressure_profiles_parametric}, the high accuracy of the model, at different $\alpha$ is confirmed. Fig.~(d) also shows the prediction of the non-parametric model V for case~B, matching the parametric model perfectly.\par
\begin{figure}[th]
    \centering
    \includegraphics[width=0.6\linewidth]{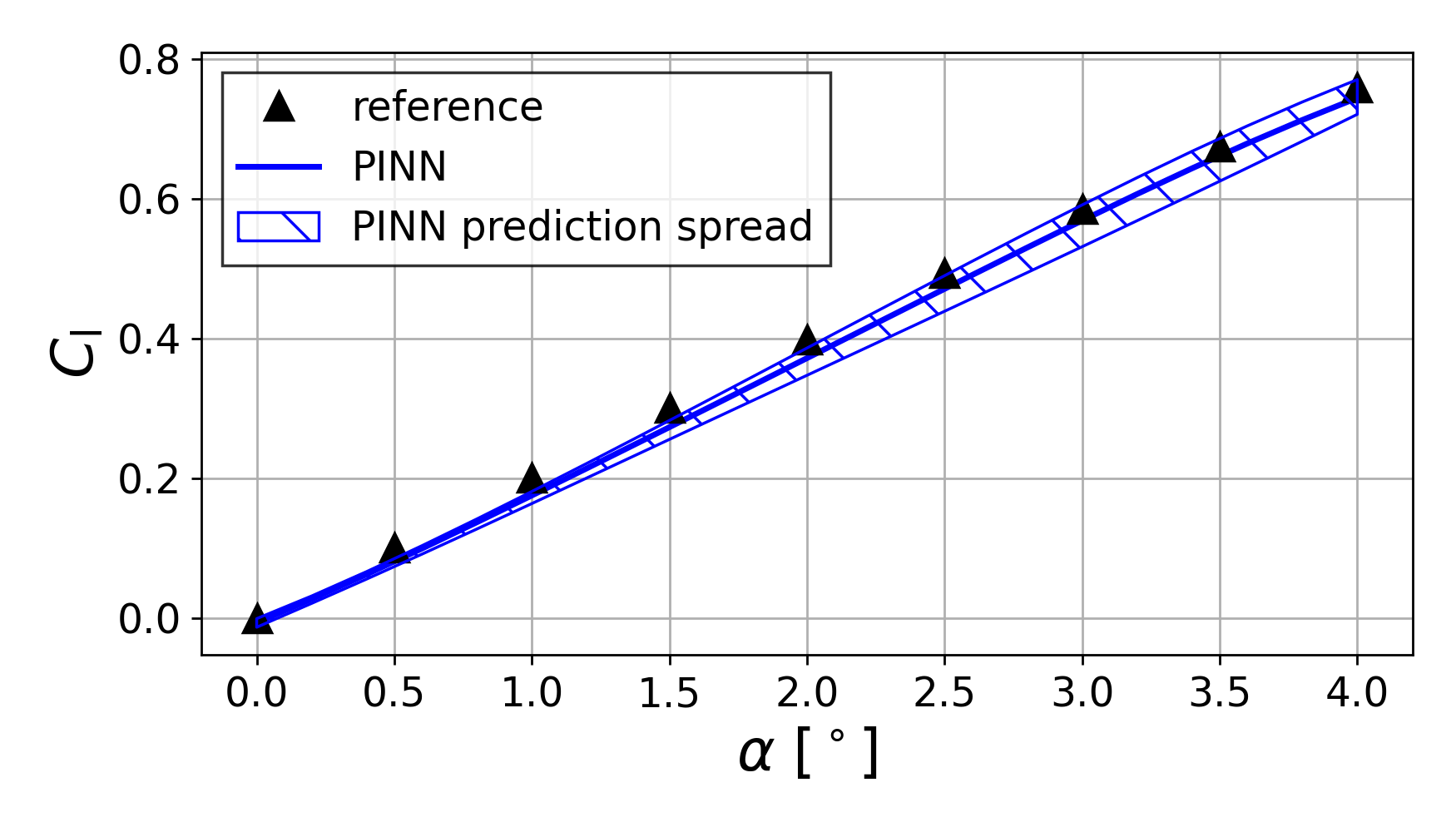}
    \caption{Predicted coefficient of lift for different angles of attack.}
    \label{fig:Cl_plot}
\end{figure}

In Fig.~\ref{fig:Cl_plot}, the mean coefficient of lift prediction over four randomly initialized training runs is shown. For the entire parameter range, the lift agrees well with the reference. The hatched area shows the minimum and maximum predictions over four training runs. Overall, the model performs quite consistently with almost identical lifts at lower $\alpha$ and slightly larger spread in the transonic region.\par
The quantitative accuracy of the predictions is again evaluated using the $RMAE$, $MAE$ and $R_2$-score over four randomly initialized training runs, as explained in Sec.~\ref{sec:results_baseline}. The error metrics are summarized in Tab~\ref{tab:errors_parametric}. Similarly to the non-parametric version, we see errors below one percent for all angles of attack. Overall, we can see that the proposed PINN methodology provides good estimations of the lifting coefficient for the entire parameter range. The results do, however, also reveal a current limitation of the method. More specifically, very weak shocks are not accurately captured because the sensor is too dissipative for these scenarios. Possible future improvements of the method to overcome this limitation are discussed in Sec,~\ref{sec:conclusion_and_outlook}. The training time varies between $1.5\;\text{h}$ and $2.75\;\text{h}$ due to the convergence criteria in the inner L-BFGS iterations. We have however observed that the accuracy does not significantly improve anymore after one epoch, which requires about $30\;\min$ on average. Predicting a specific solution at an angle of attack of interest after the training is possible in about $35\,\mathrm{ms}$.
\begin{table}[tb]
    \centering
    \caption{Mean percentage errors for parametric PINN at $M_\infty=0.72$ for different $\alpha$.}
    \label{tab:errors_parametric}
    \begin{tabular}{|c||c|c|c|c|c|c|}
                \hline
                $\alpha~[^\circ]$
                & \multicolumn{2}{c|}{$\bm{RMAE}\;[\%]$} &
                  \multicolumn{2}{c|}{$\bm{MAE}$} &
                  \multicolumn{2}{c|}{$\bm{R_2}\cdot 10$}\\
                  \hline
                  \hline
                0 &
                \makecell[l]{$C_p$ \\ $M$ } & 
                \makecell[c]{$0.12\pm 0.04$ \\ $0.1\pm 0.04$} &
                \makecell[l]{$C_p$ \\ $M$} & 
                \makecell[c]{$(1.8\pm 0.6)\cdot 10^{-3}$ \\ $(7\pm 2.5)\cdot 10 ^{-4}$} &
                \makecell[l]{$C_p$ \\ $M$} & 
                \makecell[c]{$9.996\pm 0.003$ \\ $9.994\pm 0.003$}\\
                \hline
                1 &
                \makecell[l]{$C_p$ \\ $M$ } & 
                \makecell[c]{$0.3\pm 0.08$ \\ $0.27\pm 0.07$} &
                \makecell[l]{$C_p$ \\ $M$} & 
                \makecell[c]{$(5.9\pm 1.5)\cdot 10^{-3}$ \\ $(2.3\pm 0.6)\cdot 10 ^{-3}$} &
                \makecell[l]{$C_p$ \\ $M$} & 
                \makecell[c]{$9.96\pm 0.012$ \\ $9.958\pm 0.012$}\\
                \hline
                2 &
                \makecell[l]{$C_p$ \\ $M$ } & 
                \makecell[c]{$0.32\pm 0.13$ \\ $0.29\pm 0.11$} &
                \makecell[l]{$C_p$ \\ $M$} & 
                \makecell[c]{$(8\pm 4)\cdot 10^{-3}$ \\ $(3.2\pm 1.2)\cdot 10 ^{-3}$} &
                \makecell[l]{$C_p$ \\ $M$} & 
                \makecell[c]{$9.88\pm 0.04$ \\ $9.85\pm 0.07$}\\
                \hline
                3 &
                \makecell[l]{$C_p$ \\ $M$ } & 
                \makecell[c]{$0.32\pm 0.14$ \\ $0.33\pm 0.1$} &
                \makecell[l]{$C_p$ \\ $M$} & 
                \makecell[c]{$(9\pm 4)\cdot 10^{-3}$ \\ $(4.3\pm 1.3)\cdot 10 ^{-3}$} &
                \makecell[l]{$C_p$ \\ $M$} & 
                \makecell[c]{$9.88\pm 0.06$ \\ $9.84\pm 0.08$}\\
                \hline
                4 &
                \makecell[l]{$C_p$ \\ $M$ } & 
                \makecell[c]{$0.44\pm 0.06$ \\ $0.47\pm 0.04$} &
                \makecell[l]{$C_p$ \\ $M$} & 
                \makecell[c]{$(1.24\pm 0.17)\cdot 10^{-2}$ \\ $(6.73\pm 0.5)\cdot 10 ^{-3}$} &
                \makecell[l]{$C_p$ \\ $M$} & 
                \makecell[c]{$9.88\pm 0.03$ \\ $9.831\pm 0.022$}\\
                \hline
    \end{tabular}
\end{table}

\section{Conclusion}
\label{sec:conclusion_and_outlook}
In this work, we combine a novel artificial viscosity based PINN model with mesh transformations enabling, for the first time, the simulation of transonic flows around airfoils with a PINN based solver. Normal shock waves are identified with a newly introduced analytical sensor function, based on pressure gradients in flow direction. The sensor locally applies additional AV, which stabilizes these shocks during training. Three different types of AV are proposed and tested on the NACA~0012 airfoil at different transonic flow conditions. The combination of the matrix-valued AV with the shock and the stagnation point sensor performs best, consistently achieving relative errors of less than $1~\%$, compared to reference finite volume solutions, except for the highest Mach number case. In addition, we demonstrate the usage of the model on a parametric test-case where a single neural network is trained in a continuous range of angle of attacks of $\alpha\in(0, 4)$ at $M=0.72$ achieving similar accuracies as the non-parametric model. The training times range from $30-45\,\mathrm{min}$ for the non-parametric, to $90-165\,\mathrm{min}$ for the parametric model. \par
Overall, the methodology accurately predicts flow fields and hence also yields accurate lift coefficients. It significantly improves predictions of mesh-transformation based PINNs in the transonic regime. To the authors knowledge, this is the first PINN-based method that can provide accurate predictions for transonic flows around airfoils without using any supervised terms.
A notable takeaway is that the removal of viscosity from the stagnation points is crucial to obtain accurate predictions (see \ref{fig:pressure_field_baseline}~(b)-(c)). We suspect that an inappropriate amount of viscosity at the leading edge stagnation point leads to inaccuracies which are propagated downstream along the rest of the surface. Hence, the prediction accuracy in the whole flow field is affected. The proposed AV, thus requires the removal of these deficiencies.\par 
In the following, we discuss current limitations of the method and possible directions for future improvements. Firstly, as seen in Fig.~\ref{fig:surface_Cp_all_cases}, when applying the method to a case which combines a high Mach numbers with a higher angle of attack, the shock position is shifted upstream. Interestingly, when using ADAM~\cite{Kingma.22.12.2014}, instead of L-BFGS, as the optimizer, we observed that the shock moves too far downstream during training. Hence, we suspect that the issue is related to the choice of the optimization algorithm. Recently, Cao~and~Zhang~\cite{Cao.2025} have analyzed the ill-conditioning of neural networks and proposed time-stepping-oriented neural networks to alleviate the ill-conditioning of the underlying system when training PINNs. They propose a quasi-time stepping optimization procedure to alleviate the ill-conditioning. Future extensions of the presented method could make use of such problem-specific optimization procedures to improve the prediction accuracy, even for higher Mach numbers. 
Secondly, we observed that the shocks are less sharp than for the reference solutions. This increased dissipativeness can also be confirmed using the total pressure loss, as analyzed in~\ref{sec:pressure_loss}. Comparing the locations where the sensor is active (in Fig.~\ref{fig:sensor_field_baseline}) to the sensor functions in legacy methods (e.g. \cite[p.50]{Langer.2022b}), we see that the active area, is much larger. This possibly contributes to the higher pressure losses. However, empirical tests have also shown that an increase in $k_\mathrm{shock}^{(0)}$ does not lead to higher accuracies and can potentially reduce the robustness. Another possibility to reduce dissipativeness, would be to perform mesh refinement around the shock location. So-called h-refinement is well known for classical CFD methods~\cite[p. 304]{Blazek.2015} and could potentially help to obtain more accurate approximations of the shock. Another possible solution would be to let the NN itself predict the viscosity distribution, as for example shown in~\cite{Wassing.2024, Wagenaar.2023} to ensure that only the necessary amount of AV is applied. Here, the main challenge would be to find a training procedure that can reliably produce an adequate distribution for transonic problems. We use an analytical sensor instead, since it has shown to converge more consistently and provides more control on where the AV is applied.\par
In our experience, the accuracy of the predictions is also sensitive to the viscosity factor $\nu$. If $\nu$ is too high, the resulting shocks are not well resolved. If $\nu$ is too low, the method is, in general, inconsistent and the shock position is not accurate. We have found that values of $\nu\approx 2\cdot 10^{-3}$ are in general a good compromise and perform well on most test-cases. Hence, $\nu$ typically does not need to be set to a different value when dealing with a new test-case.\par
Overall, we conclude that PINNs are currently unable to outperform state-of-the-art CFD methods in terms of speed and accuracy for non-parametric forward problems in most scenarios. However, as recently demonstrated in the various works of Cao~et~al.~\cite{Cao.2024c, Cao.2025b, Cao.02.01.2025}, the mesh transformation based PINNs can address highly parametric engineering tasks by solving PDEs in a continuous state space. Therefore, we believe that with future improvements of the PINN approach, the method may become a useful tool for rapid exploration of large design spaces in aerodynamics. Our work addresses a crucial missing component, that is the approximation of transonic flows. Whether parametric PINN methods will be favorable, compared to traditional, data-driven parametric models needs to be decided for each specific task at hand. In the near future, the main challenges are to improve robustness and to decrease the dissipativeness of the sensor. 
\clearpage

\begin{acknowledgments}
This article has been submitted to Physics of Fluids. After it is published, it will be found at \href{https://pubs.aip.org/aip/pof}{https://pubs.aip.org/aip/pof}.\\
This project was made possible by the DLR Quantum Computing Initiative and the Federal Ministry for Economic Affairs and Climate Action; \href{https://qci.dlr.de/toquaflics/}{qci.dlr.de/projects/toquaflics}\\ The authors gratefully acknowledge the scientific support and HPC resources provided by the German Aerospace Center (DLR). The HPC system CARA is partially funded by "Saxon State Ministry for Economic Affairs, Labour and Transport" and "Federal Ministry for Economic Affairs and Climate Action".\\ All reference simulations were calculated using CODA. CODA is the computational fluid dynamics (CFD) software being developed as part of a collaboration between the French Aerospace Lab ONERA, the German Aerospace Center (DLR), Airbus, and their European research partners. CODA is jointly owned by ONERA, DLR and Airbus. 
\end{acknowledgments}

\section*{Data Availability Statement}
The data that support the findings of this study are available from the corresponding author upon reasonable request.
\clearpage
\appendix
\section{Error Metrics}
The $MAE$ and $R_2$-score for the results in Sec.~\ref{sec:results_baseline} are shown in Tabs.~\ref{tab:errors_non_parametric_MAE}-\ref{tab:errors_non_parametric_r2}
\label{sec:apx_error_metrics}
\begin{center}
    {
    \captionsetup{type=tabular}
    \captionof{table}{$MAE\cdot 10^{2}$s for different test cases and field variables $C_p$ and $M$.}
    \label{tab:errors_non_parametric_MAE}
    \begin{tabular}{|c||c|c|c|c|c|c|c|c|}
                \hline
                \rotatebox[origin=c]{90}{Model}
                & \multicolumn{2}{c|}{\makecell[c]{Case A \\$M_\infty =0.7 $\\$\alpha = 4^\circ$}} &
                  \multicolumn{2}{c|}{\makecell[c]{Case B \\$M_\infty =0.72 $\\$\alpha = 3^\circ$}} &
                  \multicolumn{2}{c|}{\makecell[c]{Case C \\$M_\infty =0.75 $\\$\alpha = 2^\circ$}} &
                  \multicolumn{2}{c|}{\makecell[c]{Case D \\$M_\infty =0.78 $\\$\alpha = 3^\circ$}}\\ \hline \hline
                0 &
                \makecell[l]{$C_p$ \\ $M$ } & 
                \makecell[c]{$13.6\pm 0.7$ \\ $5.19\pm 2.6$} &
                \makecell[l]{$C_p$ \\ $M$} & 
                \makecell[c]{$10.7\pm 0.8$ \\ $4.37\pm 0.29$} &
                \makecell[l]{$C_p$ \\ $M$} & 
                \makecell[c]{$8.8\pm 0.24$ \\ $3.8\pm 1$} &
                \makecell[l]{$C_p$ \\ $M$} & 
                \makecell[c]{$22.2\pm 0.4$ \\ $9.43\pm 0.15$}\\
                \hline
                I &
                \makecell[l]{$C_p$ \\ $M$} & 
                \makecell[c]{$6\pm1$ \\ $2.4\pm0.4$} &
                \makecell[l]{$C_p$ \\ $M$} & 
                \makecell[c]{$11\pm7$ \\ $23\pm25$} &
                \makecell[l]{$C_p$ \\ $M$} & 
                \makecell[c]{$4.4\pm0.5$ \\ $1.92\pm0.19$} &
                \makecell[l]{$C_p$ \\ $M$} & 
                \makecell[c]{$13\pm5$ \\ $21\pm30$}\\
                \hline
                II &
                \makecell[l]{$C_p$ \\ $M$} & 
                \makecell[c]{$1.35\pm0.07$ \\ $0.639\pm0.024$} &
                \makecell[l]{$C_p$ \\ $M$} & 
                \makecell[c]{$\mathit{15\pm18}$ \\ $\mathit{19\pm24}$} &
                \makecell[l]{$C_p$ \\ $M$} & 
                \makecell[c]{$2.2\pm2$ \\ $1\pm 0.9$} &
                \makecell[l]{$C_p$ \\ $M$} & 
                \makecell[c]{$\mathbf{4.7\pm0.4}$ \\ $\mathbf{2.3\pm0.17}$}\\
                \hline
                III &
                \makecell[l]{$C_p$ \\ $M$} & 
                \makecell[c]{$\mathit{6\pm8}$ \\ $\mathit{6\pm11}$} &
                \makecell[l]{$C_p$ \\ $M$} & 
                \makecell[c]{$\mathit{6\pm9}$ \\ $\mathit{13\pm24}$} &
                \makecell[l]{$C_p$ \\ $M$} & 
                \makecell[c]{$1.19\pm0.09$ \\ $0.57\pm0.04$} &
                \makecell[l]{$C_p$ \\ $M$} & 
                \makecell[c]{$12\pm9$ \\ $18\pm28$}\\
                \hline
                IV &
                \makecell[l]{$C_p$ \\ $M$} & 
                \makecell[c]{$1.4\pm0.4$ \\ $0.66\pm0.15$} &
                \makecell[l]{$C_p$ \\ $M$} & 
                \makecell[c]{$0.91\pm0.29$ \\ $0.45\pm0.12$} &
                \makecell[l]{$C_p$ \\ $M$} & 
                \makecell[c]{$1.4\pm0.4$ \\ $0.65\pm0.15$} &
                \makecell[l]{$C_p$ \\ $M$} & 
                \makecell[c]{$7.9\pm0.4$ \\ $3.91\pm0.15$}\\
                \hline
                V &
                \makecell[l]{$C_p$ \\ $M$} & 
                \makecell[c]{$\mathbf{1.17\pm0.09}$ \\ $\mathbf{0.58\pm 0.03}$} &
                \makecell[l]{$C_p$ \\ $M$} & 
                \makecell[c]{$\mathbf{0.78\pm0.19}$ \\ $\mathbf{0.41\pm0.07}$} &
                \makecell[l]{$C_p$ \\ $M$} & 
                \makecell[c]{$\mathbf{0.79\pm0.27}$ \\ $\mathbf{0.42\pm0.11}$} &
                \makecell[l]{$C_p$ \\ $M$} & 
                \makecell[c]{$7.5\pm1$ \\ $3.7\pm0.5$}\\
                \hline
    \end{tabular}}
\end{center}
\clearpage
\begin{center}
    {\captionsetup{type=tabular}
    \captionof{table}{$R_2\cdot 10$ for different test cases and field variables.}
    \label{tab:errors_non_parametric_r2}
    \begin{tabular}{|c||c|c|c|c|c|c|c|c|}
                \hline
                  \rotatebox[origin=c]{90}{Model}
                 &\multicolumn{2}{c|}{\makecell[c]{Case A \\$M_\infty =0.7 $\\$\alpha = 4^\circ$}} &
                  \multicolumn{2}{c|}{\makecell[c]{Case B \\$M_\infty =0.72 $\\$\alpha = 3^\circ$}} &
                  \multicolumn{2}{c|}{\makecell[c]{Case C \\$M_\infty =0.75 $\\$\alpha = 2^\circ$}} &
                  \multicolumn{2}{c|}{\makecell[c]{Case D \\$M_\infty =0.78 $\\$\alpha = 3^\circ$}}\\ \hline \hline
                 0 &
                \makecell[l]{$C_p$ \\ $M$ } & 
                \makecell[c]{$5.2\pm 0.4$ \\ $5.2\pm 0.3$} &
                \makecell[l]{$C_p$ \\ $M$} & 
                \makecell[c]{$5.8\pm 0.4$ \\ $5.7\pm 0.4$} &
                \makecell[l]{$C_p$ \\ $M$} & 
                \makecell[c]{$6.2\pm 1.1$ \\ $6.1\pm 1.1$} &
                \makecell[l]{$C_p$ \\ $M$} & 
                \makecell[c]{$1\pm 0.18$\\ $1.34\pm 0.15$}\\
                \hline
                 I &
                \makecell[l]{$C_p$ \\ $M$ } & 
                \makecell[c]{$8\pm 0.5$ \\ $7.7\pm 0.5$} &
                \makecell[l]{$C_p$ \\ $M$} & 
                \makecell[c]{$4\pm 6$ \\ $-1\pm 14$} &
                \makecell[l]{$C_p$ \\ $M$} & 
                \makecell[c]{$8.1\pm 0.4$ \\ $7.8\pm 0.5$} &
                \makecell[l]{$C_p$ \\ $M$} & 
                \makecell[c]{$3\pm 5$ \\ $-38\pm 9$}\\
                \hline
                 II &
                \makecell[l]{$C_p$ \\ $M$ } & 
                \makecell[c]{$9.85\pm 0.04$ \\ $9.79\pm 0.06$} &
                \makecell[l]{$C_p$ \\ $M$} & 
                \makecell[c]{$\mathit{0\pm 14}$ \\ $\mathit{-76\pm 120}$} &
                \makecell[l]{$C_p$ \\ $M$} & 
                \makecell[c]{$9.3 \pm 1$ \\ $9.2\pm 1$} &
                \makecell[l]{$C_p$ \\ $M$} & 
                \makecell[c]{$\mathbf{8.56\pm 0.15}$ \\ $\mathbf{8.32\pm 0.15}$}\\
                \hline
                 III &
                \makecell[l]{$C_p$ \\ $M$ } & 
                \makecell[c]{$\mathit{8\pm 4}$ \\ $\mathit{-8\pm 34}$} &
                \makecell[l]{$C_p$ \\ $M$} & 
                \makecell[c]{$\mathit{7\pm 6}$ \\ $\mathit{-8\pm 16}$} &
                \makecell[l]{$C_p$ \\ $M$} & 
                \makecell[c]{$9.81\pm 0.14$ \\ $9.76\pm 0.19$} &
                \makecell[l]{$C_p$ \\ $M$} & 
                \makecell[c]{$5\pm 5$ \\ $-3\pm 8$}\\
                \hline
                 IV &
                \makecell[l]{$C_p$ \\ $M$ } & 
                \makecell[c]{$9.84\pm 0.05$ \\ $9.77\pm 0.08$} &
                \makecell[l]{$C_p$ \\ $M$} & 
                \makecell[c]{$9.86\pm 0.04$ \\ $9.82\pm 0.06$} &
                \makecell[l]{$C_p$ \\ $M$} & 
                \makecell[c]{$9.75\pm 0.1$ \\ $9.67\pm $0.12} &
                \makecell[l]{$C_p$ \\ $M$} & 
                \makecell[c]{$7.27\pm 0.17$ \\ $6.92\pm $0.18}\\
                \hline
                 V &
                \makecell[l]{$C_p$ \\ $M$ } & 
                \makecell[c]{$\mathbf{9.88\pm 0.01}$ \\ $\mathbf{9.819\pm 0.011}$} &
                \makecell[l]{$C_p$ \\ $M$} & 
                \makecell[c]{$\mathbf{9.89\pm 0.02}$ \\ $\mathbf{9.85\pm 0.03}$} &
                \makecell[l]{$C_p$ \\ $M$} & 
                \makecell[c]{$\mathbf{9.86\pm 0.04}$ \\ $\mathbf{9.83\pm 0.06}$} &
                \makecell[l]{$C_p$ \\ $M$} & 
                \makecell[c]{$7.4\pm 0.4$ \\ $7.1\pm 0.5$}\\
                \hline
    \end{tabular}}
\end{center}

\section{Reference Simulations}
\label{sec:reference_simulation_coda}
For all reference simulations, we use CODA. CODA is the computational fluid dynamics (CFD) software being developed as part of a collaboration between the French Aerospace Lab ONERA, the German Aerospace Center (DLR), Airbus, and their European research partners. CODA is jointly owned by ONERA, DLR and Airbus. The used o-type grid has a total of $400\times 96$ points with $400$ points on the airfoil's surface. The outer boundary of the domain is located $90$ chord lengths away from the airfoil. The accuracy has been confirmed with a mesh convergence study. Note that this grid is different to the grid used for the PINN model~(see Fig.~\ref{fig:grid}). All residuals are converged to an order $10^{-12}$. We use a second order finite-volume discretization. No far-field correction (see e.g.~\cite[pp.262-268]{Blazek.2015}) has been used. The convective fluxes are evaluated using Roe's approximate Riemann solver~\cite{Roe.1997}. For the quantitative comparisons in Tabs.~\ref{tab:errors_non_parametric_RMAE}--\ref{tab:errors_non_parametric_r2} we interpolate the PINN prediction and the finite volume reference to a Cartesian grid in the physical domain. Points inside the airfoil are excluded. We then calculate the errors based on the points in this regular grid. 
\section{Construction of Matrix-Valued Artificial Viscosity}
\label{sec:apx_matrix_diss}
The Matrices $|A|$ and $|B|$ are constructed as discussed in~\cite{Swanson.1992}. However, since we evaluate Eq.~\eqref{eq:loss_residual} in Cartesian coordinates, the formulation simplifies with $\xi=x, \eta=y, J=1$. Hence, the eigenvalues for $A$ are given by:
\begin{equation}
    \begin{aligned}
        \lambda_1 &= u + c\\
        \lambda_2 &= u - c\\
        \lambda_3 &= u\\
    \end{aligned}
\end{equation}
As mentioned above, for the construction of $|A|$, the absolute eigenvalues are limited. The limited values are given by:
\begin{equation}
    \begin{aligned}
        |\tilde{\lambda}_1| &= \max (|\lambda_1|, \varepsilon)\\
        |\tilde{\lambda}_2| &= \max (|\lambda_3|, \varepsilon)\\
        |\tilde{\lambda}_3| &= \max (|\lambda_4|, \varepsilon),\\
    \end{aligned}
\end{equation}
with $\varepsilon=0.25$. For $|B|$, $u$ is replaced by $v$.
\section{PINN without Artificial Viscosity}
\label{sec:apx_pinn_without_av}
\begin{figure}[h]
    \centering
    \includegraphics[width=0.5\linewidth]{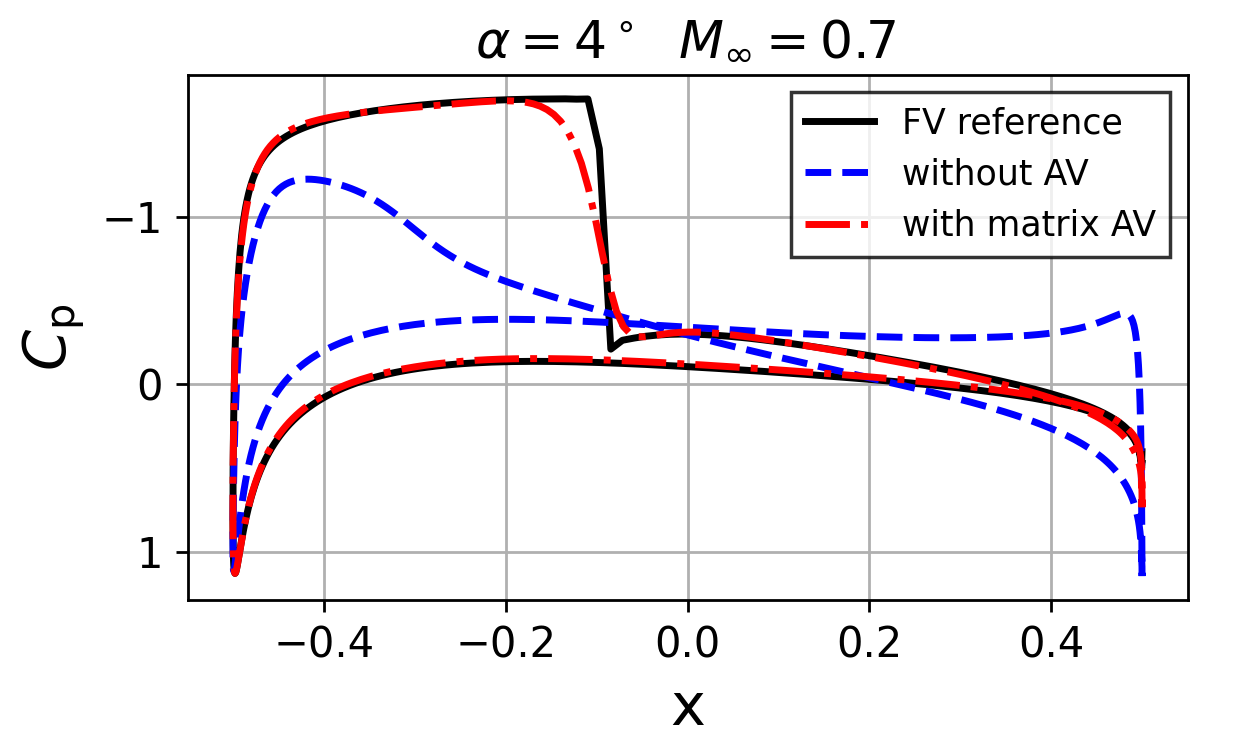}
    \caption{Comparison of PINN prediction with and without AV. Model V (see Tab.~\ref{tab:test-cases}) is used as the AV variant.}
    \label{fig:pinn_without_AV}
\end{figure}

To highlight the failure of standard PINNs to capture shock waves in transonic flows, we consider case~A in Tab.~\ref{tab:test-cases}. Fig.~\ref{fig:pinn_without_AV}~shows predicted surface pressure, when using the full PINN model with MT and identical hyperparameters but with no artificial viscosity. Evidently, the model fails completely to capture the correct pressure field and no approximation of the shock is visible. As described in Sec.~\ref{sec:methods_artificial_viscosity}, the shock solution is not a minimum of the standard $L_2$-norm based residual loss. Hence, the model fails. The introduction of AV smooths out the shock, enabling the approximation of the modified solution with the PINN, as shown for the matrix-valued AV.
\section{Total Pressure Loss}
\label{sec:pressure_loss}
\begin{figure}[h]
    \centering
    \includegraphics[width=.9\linewidth]{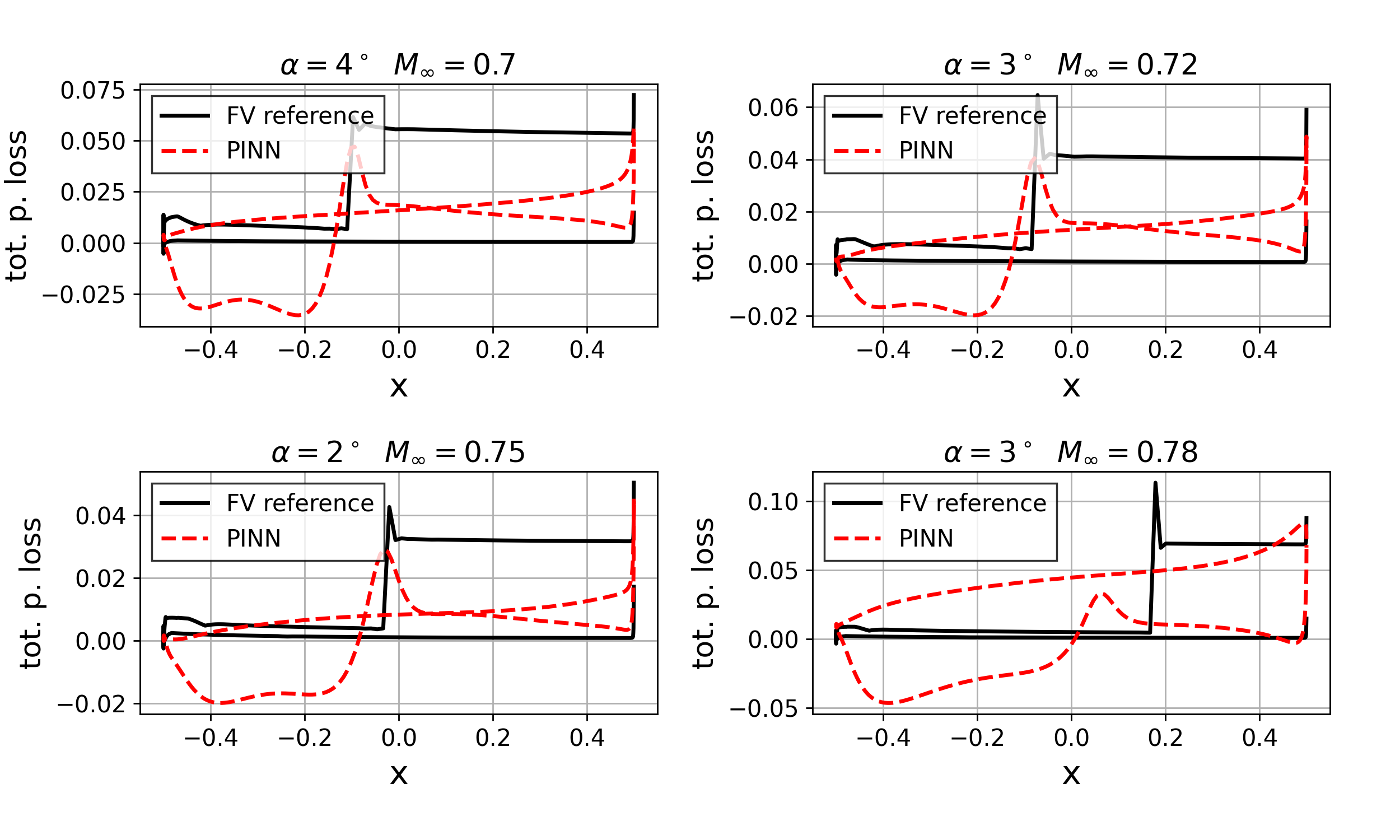}
    \caption{Total pressure loss for the non-parametric model V in Sec.~\ref{sec:results_baseline}.}
    \label{fig:pressure_loss}
\end{figure}

The total pressure is defined as:
\begin{equation}
    p_\mathrm{tot} = p \left(1+ \frac{1}{2}(\kappa-1)M^2\right)^{\dfrac{\kappa}{\kappa - 1}},
    \label{eq:tot_pressure}
\end{equation}
with the local mach number $M=\|\bm{q}\|/c$. The far field total pressure $p_{\mathrm{tot}, \infty}$ is obtained when evaluating Eq.~\eqref{eq:tot_pressure} at the far field conditions, given by $\bm{w_\infty}= (\rho_\infty, u_\infty, v_\infty, p_\infty)$. The total pressure loss is then given by:
\begin{equation}
 C_{p_\mathrm{tot}, \mathrm{loss}} = 1 - \frac{p_\mathrm{tot}}{p_{\mathrm{tot}, \infty}}.
 \label{eq:p_loss}
\end{equation}
The total pressure loss on the airfoil's surface can serve as a measure of the accuracy of the numerical scheme. Fig.~\ref{fig:pressure_loss} shows the pressure loss for the PINN model and the reference simulation for the four test-cases in Sec.~\ref{sec:results_baseline}. As highlighted by the reference simulation, we generally expect pressure losses close to $0$ for subsonic flows, a steep increase around shocks and a constant, increased value after shocks. For the PINN, the increase around the shock is on the same order of magnitude as for the reference. We do however see significant changes in the pressure loss in regions besides the shock, which indicates, that the PINN methodology is more viscous and further improvements can be made, as discussed in Sec.~\ref{sec:conclusion_and_outlook}.
\section{Reconstruction of Derivatives using Inverse Metric Terms}
\label{sec:derivatives_in_mesh_trafo}
The inverse metric terms are the derivatives of $x$ and $y$ with respect to the computational coordinates $\xi$ and $\eta$. Given the discrete mapping in Eq.~\eqref{eq:discrete_mesh_trafo}, they can be approximated with finite differences. Using the inverse mesh metric terms, the derivatives of a function $f$ with respect to the physical coordinates $x$ and $y$ can then be reconstructed:
{\small
    \begin{equation}
        \begin{aligned}
            J &= \det \begin{pmatrix}
                \pdv{x}{\xi}, &\pdv{y}{\xi}\\
                \pdv{x}{\eta}, &\pdv{y}{\eta}
            \end{pmatrix}\\
            \pdv{f}{x} &= \dfrac{1}{J}\left[ \pdv{f}{\xi} \pdv{y}{\eta} - \pdv{f}{\eta}\pdv{y}{\xi}\right],\\
            \pdv{f}{y} &= \dfrac{1}{J}\left[ \pdv{f}{\eta} \pdv{x}{\xi} - \pdv{f}{\xi}\pdv{x}{\eta}\right],\\
            \pdv[2]{f}{x} &= \dfrac{1}{J^2} \left[ \pdv[2]{f}{\xi} \left( \pdv{y}{\eta}\right)^2 - 2 \pdv{f}{\xi}{\eta}\pdv{y}{\xi}\pdv{y}{\eta} + \pdv[2]{f}{\eta} \left( \pdv{y}{\xi}\right)^2 \right] +\\
            & \left[ \left(\pdv{y}{\eta}\right)^2 \pdv[2]{y}{\xi} - 2 \pdv{y}{\xi}\pdv{y}{\eta}\pdv{y}{\xi}{\eta} + \left( \pdv{y}{\xi}\right)^2\pdv[2]{y}{\eta}\right] \left[ \pdv{f}{\xi}\pdv{x}{\eta} - \pdv{f}{\eta}\pdv{x}{\xi}\right] +\\
            \dfrac{1}{J^3} &\left[ \left( \pdv{y}{\eta}\right)^2\pdv[2]{x}{\xi} -2\pdv{y}{\xi}\pdv{y}{\eta}\pdv{x}{\xi}{\eta} + \left(\pdv{y}{\xi}\right)^2\pdv[2]{x}{\eta}\right]\left[\pdv{f}{\eta}\pdv{y}{\xi} - \pdv{f}{\xi}\pdv{y}{\eta}\right]\\
            \pdv[2]{f}{y} &= \dfrac{1}{J^2} \left[ \pdv[2]{f}{\xi} \left( \pdv{x}{\eta}\right)^2 - 2 \pdv{f}{\xi}{\eta}\pdv{x}{\xi}\pdv{x}{\eta} + \pdv[2]{f}{\eta} \left( \pdv{x}{\xi}\right)^2 \right] +\\
            & \left[ \left(\pdv{x}{\eta}\right)^2 \pdv[2]{y}{\xi} - 2 \pdv{x}{\xi}\pdv{x}{\eta}\pdv{y}{\xi}{\eta} + \left( \pdv{x}{\xi}\right)^2\pdv[2]{y}{\eta}\right] \left[ \pdv{f}{\xi}\pdv{x}{\eta} - \pdv{f}{\eta}\pdv{x}{\xi}\right] +\\
            \dfrac{1}{J^3} &\left[ \left( \pdv{x}{\eta}\right)^2\pdv[2]{x}{\xi} -2\pdv{x}{\xi}\pdv{x}{\eta}\pdv{x}{\xi}{\eta} + \left(\pdv{x}{\xi}\right)^2\pdv[2]{x}{\eta}\right]\left[\pdv{f}{\eta}\pdv{y}{\xi} - \pdv{f}{\xi}\pdv{y}{\eta}\right].\\
        \end{aligned}
        \label{eq:apx_mesh_metrics}
    \end{equation}
}
For a derivation of these relations, the interested reader is referred to~\cite[pp. 178--183]{Anderson.1995}~\cite{MatthewMcCoy.1980}.
\clearpage
\section{Stagnation Point Sensor}
\begin{figure}[ht]
    \centering
    \includegraphics[width=0.5\linewidth]{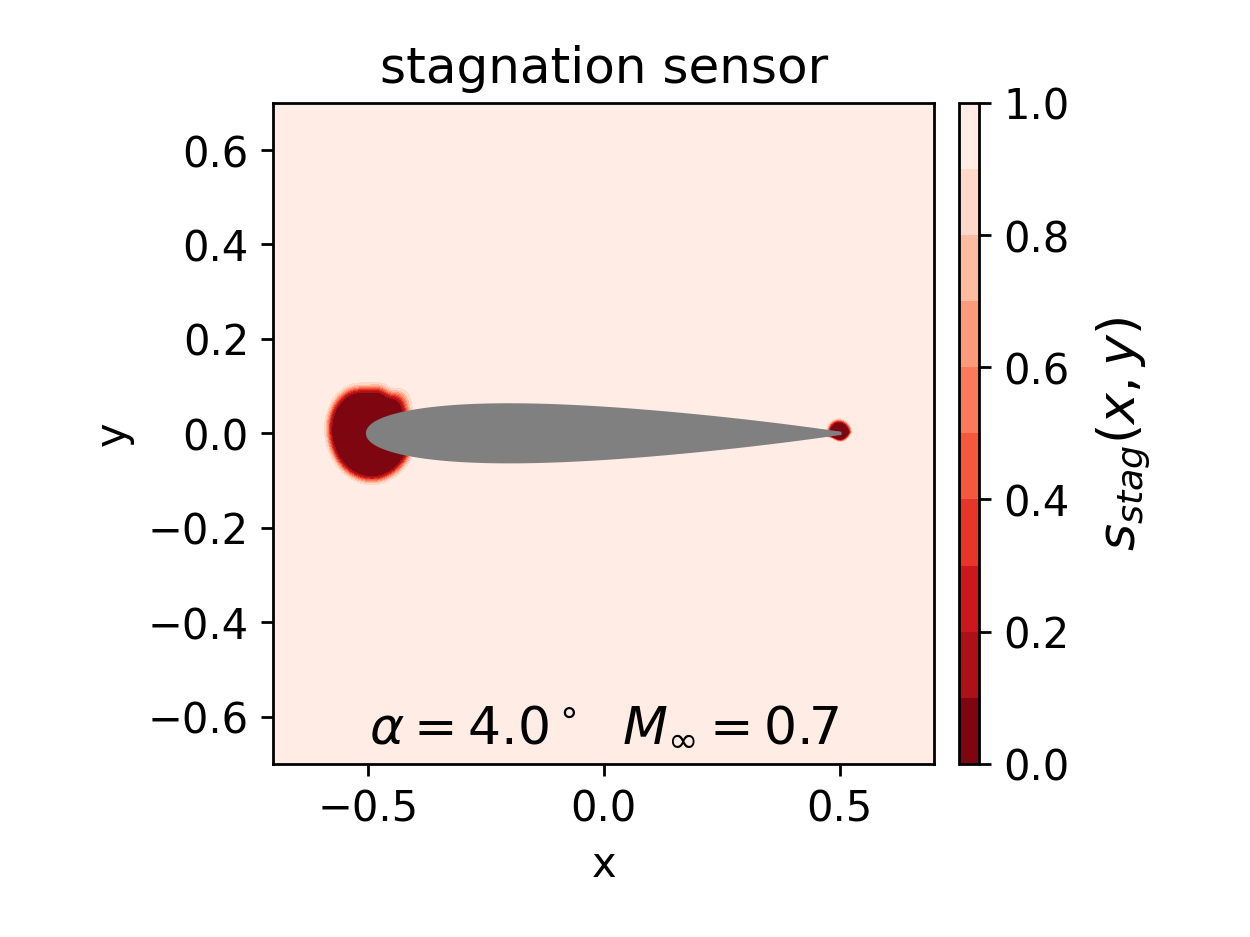}
    \caption{Stagnation point sensor for model~V and case~A.}
    \label{fig:stag_sensor}
\end{figure}

The stagnation point sensor, given by Eq.~\eqref{eq:sensor_stagnation}, can identify the stagnation points in the flow field. We observe that these should be excluded from a sensor function to facilitate the convergence of the model. For the subsonic test-case~A, the stagnation sensor $s_\mathrm{stag}$ is shown in Fig.~\ref{fig:stag_sensor}.
\clearpage

\clearpage
\clearpage
\clearpage

\newpage
\section{Hyperparameters}
\label{sec:appendix_hyperparameters}
A summary of the used hyperparameters in Sec.~\ref{sec:results_baseline} and Sec.~\ref{sec:results_parametric} is shown in Tab.~\ref{tab:training_parameters_baseline}.

\vspace{2cm}
\begin{center}
	\captionsetup{type=tabular}
	\captionof{table}{Summary of model hyperparameters in Secs.~\ref{sec:results_baseline}-\ref{sec:results_parametric}}
	\label{tab:training_parameters_baseline}
	\centering
	\begin{tabular}{|l|c|c|}
        \hline
         & non-parametric & parametric\\
        \hline \hline
        hidden layers $D$ & $5$  & $5$\\
        \hline
        neurons/layer $N_k$ & $128\;\forall\; i=1\dots D\;$ & $128\;\forall\; i=1\dots D\;$ \\
        \hline
        \makecell[l]{Fourier embedding\\frequencies $D_\mathrm{fe}$}& $128$ & $128$ \\
        \hline
        \makecell[l]{Fourier embedding\\ standard deviation $\sigma_\phi$}& $1$ & $1$\\
        \hline  
        batch size & $5000$ & $8000$\\
        \hline
        loss weighting factor residual $\lambda_\mathrm{res}$ & $2\cdot10^{4}$ & $2\cdot10^{4}$\\
        \hline
        $k_{\mathrm{stag}}^{(0)}$ & $3$ & $3$\\
        \hline
        $k_{\mathrm{stag}}^{(1)}$ & $5$ & $5$\\
        \hline
        $k_{\mathrm{shock}}^{(0)}$ & $4$ & $4$\\
        \hline
        $k_{\mathrm{stag}}^{(1)}$ &  $0.03$ & $0.03$\\
        \hline
        epochs & $40$ & $5$\\
        \hline
        L-BFGS (PyTorch): & &\\
        \texttt{max\_iter} & $1000$ & $1000$\\
        \texttt{max\_eval} & $1250$ & $1250$\\
        \texttt{tolerance\_grad} & $10^{-6}$, & $10^{-6}$\\
        \texttt{tolerance\_change} & $10^{-10}$ & $10^{-10}$\\
        \texttt{history\_size} & $1000$ & $1000$\\
        \texttt{line\_search\_fn} & \texttt{"strong\_wolfe"} & \texttt{"strong\_wolfe"}\\
        \hline
        $N_\alpha$ & / &$56\cdot 10^4$\\
        \hline
        $N_{\alpha, \mathrm{bound}}$ & / &$4\cdot 10^4$\\
        \hline
	\end{tabular}
\end{center}
\clearpage

\clearpage
\nocite{*}
\bibliography{Literatur.bib}

\end{document}